\documentclass[manuscript, nonacm]{acmart}

\AtBeginDocument{%
  }

\usepackage{tikz}
\usepackage{tcolorbox}
\usepackage{graphicx} 
\usepackage{subcaption}
\usepackage{soul}
\usepackage{tabularx}
\usepackage{array}
\usepackage{xltabular}
\usepackage[pdftex]{changebar}
\usepackage{cleveref}

\usepackage{wrapfig}
\usepackage{xspace}
\usepackage{lipsum}
\usepackage{todonotes}
\usepackage{xcolor}
\usepackage{algorithm,algpseudocode}
\usepackage{amsmath}
\usepackage{bbm}
\usepackage{wrapfig}
\usepackage{booktabs,multirow,array}
\usepackage{colortbl}
\usepackage{makecell} 
\usepackage{fontawesome5}

\definecolor{colorConfigurationLogic}{HTML}{f39882}
\definecolor{colorWorldModel}{HTML}{fddbaa}
\definecolor{colorObservation}{HTML}{d2e7ca}
\definecolor{colorCommunication}{HTML}{9ed1ea}
\definecolor{colorActions}{HTML}{e0c2dc}

\definecolor{colorEnvironment}{HTML}{7570b3}

\definecolor{colorInfrastructure}{HTML}{99BDB2}
\definecolor{colorStudyEnvironment}{HTML}{CEC6D7}
\definecolor{colorGeneral}{HTML}{DADADA}

\definecolor{colorComponentPermissionGuard}{HTML}{F6BAB0}
\definecolor{colorComponentWorldState}{HTML}{C6D38F}
\definecolor{colorComponentCommunication}{HTML}{9ED1EA}
\definecolor{colorComponentAuditLogs}{HTML}{FDDBAA}
\definecolor{colorComponentAgents}{HTML}{E1E4E4}
\definecolor{colorComponentInterface}{HTML}{FFFFFF}

\newlength\myheight
\newlength\mydepth
\settototalheight\myheight{Xygp}
\usepackage{tcolorbox}
\usepackage{tikz}

\usepackage{colortbl}
\usepackage{makecell}

\makeatletter

\newtcbox{\kwColorBox}[1][]{%
  on line, fontupper=\footnotesize\sffamily\bfseries\small,
  boxrule=0.5pt, arc=2pt, boxsep=0pt, left=1.5pt, right=1.5pt, top=1.5pt, bottom=1.5pt,
  coltext=#1!40!black, colback=#1!10!white,
  code={%
    \colorlet{tcb@temp@color}{#1}
    \extractcolorspecs{tcb@temp@color}{\tcb@temp@model}{\tcb@temp@spec}
    \convertcolorspec{\tcb@temp@model}{\tcb@temp@spec}{HTML}{\tcb@temp@html}
    \def\tcb@white@html{FFFFFF}%
    \ifx\tcb@temp@html\tcb@white@html
      \tcbset{colframe=black}%
    \else
      \tcbset{colframe=#1}%
    \fi
  }%
}

\newtcbox{\kwColorBoxRef}[1][]{%
  on line, fontupper=\footnotesize\sffamily\bfseries\small,
  boxrule=0.5pt, arc=2pt, boxsep=0pt, left=1.5pt, right=1.5pt, top=1.5pt, bottom=1.5pt,
  coltext=black, colback=#1!50!white,
  code={%
    \colorlet{tcb@temp@color}{#1}%
    \extractcolorspecs{tcb@temp@color}{\tcb@temp@model}{\tcb@temp@spec}%
    \convertcolorspec{\tcb@temp@model}{\tcb@temp@spec}{HTML}{\tcb@temp@html}%
    \def\tcb@white@html{FFFFFF}%
    \ifx\tcb@temp@html\tcb@white@html
      \tcbset{colframe=black}%
    \else
      \tcbset{colframe=#1}%
    \fi
  }%
}

\newtcbox{\kwColorBoxSpecial}[1][]{%
  on line, fontupper=\footnotesize\sffamily\bfseries\small,
  boxrule=0.5pt, arc=2pt, boxsep=0pt, left=1.5pt, right=1.5pt, top=1.5pt, bottom=1.5pt,
  colback=#1,
  code={%
    \colorlet{tcb@temp@color}{#1}%
    \extractcolorspecs{tcb@temp@color}{\tcb@temp@model}{\tcb@temp@spec}%
    \convertcolorspec{\tcb@temp@model}{\tcb@temp@spec}{HTML}{\tcb@temp@html}%
    \def\tcb@white@html{FFFFFF}%
    \ifx\tcb@temp@html\tcb@white@html
      \tcbset{colframe=black, coltext=black}%
    \else
      \tcbset{colframe=#1, coltext=white}%
    \fi
  }%
}

\newtcbox{\kwColorBoxSpecialContext}[1][]{%
  on line, fontupper=\footnotesize\sffamily\bfseries\small,
  boxrule=0.5pt, arc=2pt, boxsep=0pt, left=1.5pt, right=1.5pt, top=1.5pt, bottom=1.5pt,
  colback=#1,
  code={%
    \colorlet{tcb@temp@color}{#1}%
    \extractcolorspecs{tcb@temp@color}{\tcb@temp@model}{\tcb@temp@spec}%
    \convertcolorspec{\tcb@temp@model}{\tcb@temp@spec}{HTML}{\tcb@temp@html}%
    \def\tcb@white@html{FFFFFF}%
    \ifx\tcb@temp@html\tcb@white@html
      \tcbset{colframe=black, coltext=black}%
    \else
      \tcbset{colframe=#1, coltext=white}%
    \fi
  }%
}

\makeatother

\newcommand{\kw}[2]{%
    \begin{kwColorBox}[#2]%
    {#1}%
    \end{kwColorBox}%
}
\newcommand{\kwRef}[2]{%
    \begin{kwColorBoxRef}[#2]%
    {#1}%
    \end{kwColorBoxRef}%
}

\newcommand{\rawdesigngoaldonotuse}[2]{\kw{#1}{#2}}
\newcommand{\rawdesigngoaldonotuseRef}[2]{\kwRef{#1}{#2}}
\newcommand{\defdesigngoal}[2]{\rawdesigngoaldonotuse{\phantomsection\label{lmdesigngoal:#1}#1}{#2}}
\newcommand{\refdesigngoal}[2]{\hyperref[lmdesigngoal:#1]{\rawdesigngoaldonotuseRef{#1}{#2}}}

\newcommand{\rawcomponentdonotuse}[2]{\kwRef{#1}{#2}}
\newcommand{\rawcomponentdonotuseRef}[2]{\kwRef{#1}{#2}}
\newcommand{\defcomponent}[2]{\rawcomponentdonotuse{\phantomsection\label{lmcomponent:#1}#1}{#2}}
\newcommand{\refcomponent}[2]{\hyperref[lmcomponent:#1]{\rawcomponentdonotuseRef{#1}{#2}}}

\newcommand{\dimension}[4]{
    \vspace{-0.25em}
    \begin{tcolorbox}[
        fonttitle=\bfseries,
        coltitle=black,
        colbacktitle=#2!40,
        colback=#2!20, 
        colframe=#2,
        title={#1 \hfill}, 
        after skip=0.35em,
        left=3pt, right=3pt, top=3pt, bottom=-3pt,
        boxsep=0pt,
        toptitle=3pt, bottomtitle=3pt,
        sharp corners=all,
        boxrule=0mm, leftrule=1mm
    ]

    \end{tcolorbox}%
    \noindent
    #3
}

\usepackage{soul}
\usepackage{xcolor} %

\newcommand{\coloredul}[2]{%
  \setuldepth{yg}%
  \setul{}{2pt}%
  \setulcolor{#1}%
  \ul{#2}%
}

\usepackage{mathpazo}          %
\usepackage{xcolor}
\usepackage{listings}
\usepackage{parskip}
\usepackage{enumitem}

\definecolor{mint}{HTML}{CFEAD9}
\definecolor{forestgreen}{HTML}{1C6B45}
\definecolor{rulegray}{HTML}{A9B4A3}
\definecolor{mutedgray}{HTML}{5F6B62}

\newcommand{\Id}[1]{\text{\itshape #1}}

\begin{document}

\title[MIVAIS]{MIVAIS: A Study Environment for Multi-Agent Mixed-Initiative \\Visual Analytics Applications}

\author{Tobias Stähle}
\email{tobias.staehle@inf.ethz.ch}
\orcid{0009-0001-5983-8807}
\affiliation{%
  \institution{ETH Zürich}
  \city{Zürich}
  \country{Switzerland}
}

\author{Simon Schneider}
\email{schneiders@student.ethz.ch}
\orcid{0009-0002-5065-6799}
\affiliation{%
  \institution{ETH Zürich}
  \city{Zürich}
  \country{Switzerland}
}

\author{Rita Sevastjanova}
\email{rita.sevastjanova@inf.ethz.ch}
\orcid{0000-0002-2629-9579}
\affiliation{%
  \institution{ETH Zürich}
  \city{Zürich}
  \country{Switzerland}
}

\author{Mennatallah El-Assady}
\email{menna.elassady@inf.ethz.ch}
\orcid{0000-0001-8526-2613}
\affiliation{%
  \institution{ETH Zürich}
  \city{Zürich}
  \country{Switzerland}
}

\renewcommand{\shortauthors}{Stähle et al.}
\newcommand{\appName}{\texttt{\textbf{MIVAIS}}\xspace}

\begin{abstract}
  Mixed-initiative Visual Analytics (VA) systems empower human users by interleaving human intuition with software agents and their machine intelligence. However, the development and rigorous evaluation of such systems remain constrained by engineering overhead. Developers must, e.g., implement complex, low-level state synchronization to manage asynchronous agent behaviors, while researchers struggle to capture the multimodal provenance required to study and evaluate human-AI collaboration. We present \appName, a dual-layered research platform designed to abstract the structural complexities of mixed-initiative VA. First, it contributes a computational Infrastructure that standardizes human-software agent interaction, state synchronization, and communication between the agents. Second, it provides a declarative Study Environment that automatically logs multimodal human-AI telemetry --- including application/system state, screen capture, audio, and additional sensor data --- enabling seamless, in-situ user studies and post-session analysis. We technically validate our infrastructure by replicating three state-of-the-art systems (Podium, Voyager~2, and ProactiveVA). Furthermore, we evaluate the framework’s expressiveness and efficiency through expert case studies with HCI and VA researchers, demonstrating how \appName effectively lowers the barrier to prototyping and evaluating intelligent, co-adaptive interfaces.
\end{abstract}


\begin{teaserfigure}
 \includegraphics[width=\textwidth]{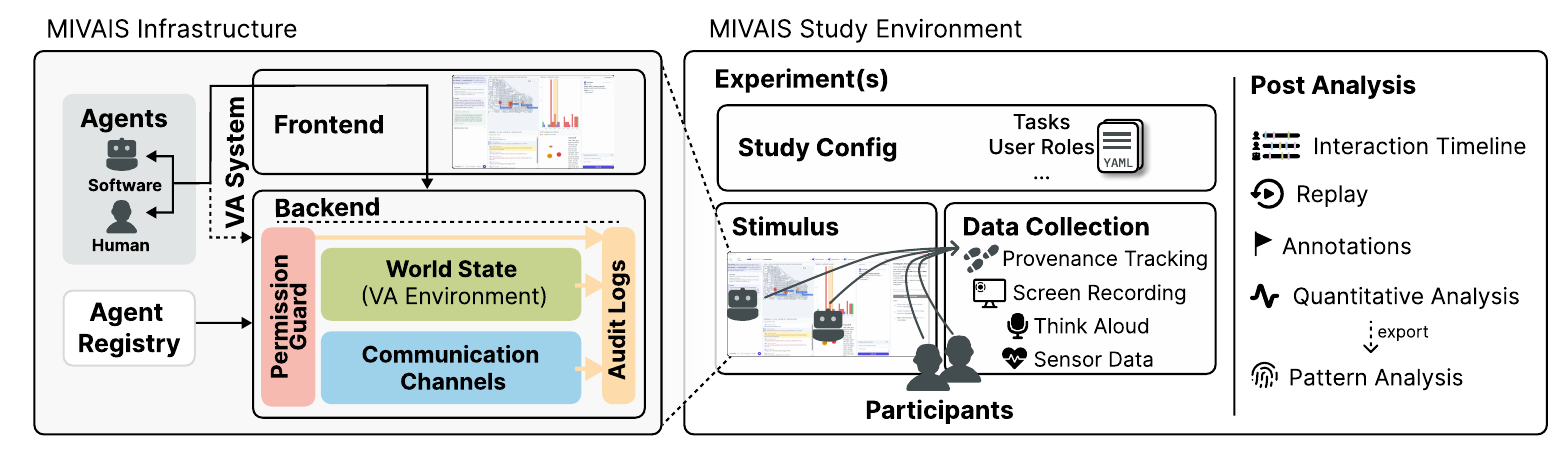}
 \vspace{-3em}
  \caption{
  Overview of the \appName framework. The \appName Infrastructure (left) supports the implementation of modular, mixed-initiative Visual Analytics (VA) systems where software and human agents collaborate. The \appName Study Environment (right) enables researchers to easily configure experiments, collect multi-modal data during participant sessions, and perform comprehensive post-analysis.
  }
  \Description{Overview of the \appName framework. The \appName Infrastructure (left) supports the implementation of modular, mixed-initiative Visual Analytics (VA) systems where software and human agents collaborate. The \appName Study Environment (right) enables researchers to easily configure experiments, collect multi-modal data during participant sessions, and perform comprehensive post-analysis.}
  \label{fig:teaser}
\end{teaserfigure}

\maketitle

\section{Introduction}

The integration of (proactive) software agents into Visual Analytics (VA) has shifted the paradigm of data analytics from human-driven querying and exploration to human-AI collaboration. In these multi-agent setups, both humans and software agents can have equal agency. In mixed-initiative VA systems, both sides share analytical tasks such as identifying data insights~\cite{dhanoa_2025_agenticVis}, generating forecasts~\cite{sivaraman_2025_divisi}, recommending visualizations~\cite{wu_2022_multiVision}, and steering underlying models~\cite{sperrle_2021_coadaptive,endert_2014_humanIsTheLoop}, while taking into account the intents of other stakeholders~\cite{Horvitz_1999_mixedInitiative}. 

Mixed-initiative and multi-agent VA systems consist of data visualizations, interface interactions, and communication among agents. Rapid progress in agent capabilities is driving an increase in the complexity of the architectures required to support their interactions. Thus, system developers must ensure continuous synchronization of a highly dynamic interface with asynchronous agents and resolve conflicts when human and software agents attempt to update the application state simultaneously. While theoretical taxonomies on human-AI collaboration and intelligent agents in VA are established~\cite{Monadjemi_2023_agentbasedFramework,staehle_2025_designspace}, there is a distinct lack of structural toolkits that abstract these engineering complexities. Consequently, Human-Computer Interaction (HCI) researchers spend a disproportionate amount of time building bespoke middleware rather than investigating the actual mixed-initiative and intelligent user interface phenomena, such as user trust, cognitive load, or multi-agent interplay. %

Recently, in the visualization community, there has been an effort to develop evaluation frameworks that systematize the research evaluation process for visualization research, e.g., reVISit~\cite{ding_2023_revisit,cutler_2026_revisit2}. %
However, the existing evaluation systems are tailored for single-user sessions with isolated visual stimuli and cannot capture the intricate dynamics of mixed-initiative collaboration. In multi-agent setups, human and AI agents continuously influence each other's behavior, introducing additional analytical and cognitive burdens that must be explicitly measured. Evaluation requires tracking of inter-agent communication to capture dynamic agency distribution and handovers, user trust, and co-adaptiveness, alongside agent behavior and feedback loops. 
This tracking is essential because participants' cognitive load stems not only from processing the underlying data but also from managing interactions with proactive agents.
To effectively assess this multi-agent interplay, researchers must go beyond standard HCI logging mechanisms and capture synchronized, multimodal provenance. This includes: underlying application state to understand the exact context of collaborative actions; provenance logs of all interacting agents to identify correlations and behavioral patterns; the (visual) stimulus of each agent to reveal the specific information that triggered decisions; the human agents verbalized reasoning (think-aloud audio) that exposes decision-making processes, trust, and mental models; and inter-agent communication logs that capture the actual dialogue and task delegation between agents. Capturing additional biometric sensor data can further provide insights into cognitive load and stress -- a capability that current study frameworks like reVISit do not support.

To address this research gap, we introduce \appName, a comprehensive research framework that streamlines the prototyping and evaluation of mixed-initiative VA applications. Bridging the gap between conceptual design, software realization, and conducting studies, \appName operationalizes theoretical frameworks on intelligent agents and mixed-initiative VA human-AI collaboration into a pragmatic, research-centric toolkit.

Our contribution is threefold:
\begin{itemize}
    \item[(1)] \textbf{A Mixed-Initiative Infrastructure Model:} A scalable software toolkit that allows structured implementation and configuration of agents and provides for high-level APIs for VA environment observation, interaction, state synchronization, communication channels among the agents, and action provenance, abstracting the low-level complexities of mixed-initiative VA systems.
    \item[(2)] \textbf{An Integrated Study Environment:} Along the Infrastructure Model, we provide a multimodal logging architecture that allows researchers to seamlessly configure and deploy user studies. By automatically logging all agents' observations, interactions, and communications and synchronizing these states with screen and audio captures, the environment enables deep post-hoc analysis of human-AI collaborative behaviors in mixed-initiative VA systems.
    \item[(3)] \textbf{Demonstration and Evaluation:} We demonstrate the capabilities of \appName by replicating state-of-the-art systems (Podium~\cite{wall_2018_podium}, Voyager~2~\cite{wongsuphasawat_2017_voyager2}, and ProactiveVA~\cite{zhao_2026_proactiveVA}). In addition, we assess its expressiveness through technical dimensions of programming systems. We also evaluate the effectiveness of \appName through expert case studies with 3 researchers showing how HCI and VA researchers and developers use \appName to effectively implement and evaluate novel mixed-initiative VA interfaces.
\end{itemize}

\appName is available as open-source software at \faGithub:~\href{https://github.com/ETH-IVIA-Lab/MIVAIS}{github.com/ETH-IVIA-Lab/MIVAIS} to support the research community with a modular, open-source foundation for future human-AI collaboration research.

\section{Background \& Related Work}

\subsection{Mixed-Initiative in Visual Analytics Systems}
Research in HCI explores how humans and computers can tackle complex problems~\cite{dellermann_2019_hybridIntelligence,Xu_2024_HAITeaming}, make informed decisions~\cite{LeeHAICollabDecisionMaking2021,reverberi_2022_experimentalHAICollab,Schemmer_2023_HAI_DecisionMaking}, and learn how data visualization and interactive exploration help humans derive insights~\cite{Keim_2008_VA}. The introduction of mixed-initiative interfaces~\cite{Horvitz_1999_mixedInitiative} has further advanced the domains of decision-making~\cite{Jiang_2018_MIDecisionMaking_HRTeams,Ju_2022_MIDecisionMaking} and visual analytics~\cite{WHChen_2025_Dango,Kcook_2015_MixedInitiativeVAUsingTaskDrivenRecommendations,sperrle_2021_LearningContextualizedUserPreferences,zhao_2025_LightVA,zhao_2025_Leva}. Several works by \citet{gotz_2006_InteractiveVisualSynthesisOfAnalyticalKnowledge}, \citet{Kcook_2015_MixedInitiativeVAUsingTaskDrivenRecommendations}, \citet{liu_2007_Tibor}, and \citet{wen_2007_ContextAwareAdaptiveInfromationRetrieval}, have established the mixed-initiative concept into VA applications, introducing software agents that support humans in navigating complex analytical tasks. Since then, researchers and developers have implemented a wide range of VA applications (e.g.,~\cite{Kcook_2015_MixedInitiativeVAUsingTaskDrivenRecommendations,ElAssady_2019_VAForTopicModeling,wall_2018_podium,Wongsuphasawat_2016_Voyager}) and many of these applications aim to generally balance automation with human steering. Nevertheless, all of these applications are implemented differently and differ in their structure and modularization. Further, many works do not provide their research prototypes or applications openly accessible to the research community, making it hard for other researchers to replicate system behavior and studies, and most publications do not provide enough details for substantial replication~\cite{fekete_2020_reproducability,cockburn_2020_threatsReplication}.

\subsection{Intelligent Agents and Human-AI Collaboration}
While the literature considers both human and AI agents as actors in mixed-initiative VA systems~\cite{dhanoa_2025_agenticVis,Monadjemi_2023_agentbasedFramework,monadjemi_2025_reviewinitiativeVA}, the implementation and operational dynamics of these intelligent agents vary widely.
Recent enhancements in Artificial Intelligence (AI) enable transforming these agents from passive to proactive actors, e.g.,~\cite{zhao_2026_proactiveVA,huang_2026_webseek}.
To identify different agent roles, behaviors, and agency distribution, work by \citet{Monadjemi_2023_agentbasedFramework}, and \citet{dhanoa_2025_agenticVis} make use of an agent-based conceptual model for mixed-initiative VA. \citet{holter_2024_deconstructing} note that the agent-centric view allows for capturing the complexity of mixed-initiative VA. Further, \citet{staehle_2025_designspace} characterize intelligent agents' logic and behavior in mixed-initiative VA through six dimensions (Configuration + Logic, World Model, Observations, Communication, Actions, and Infrastructure) by decomposing mixed-initiative VA systems into agents, an infrastructure layer, and the VA environment. 
The agent-centric view~\cite{Monadjemi_2023_agentbasedFramework} and the decomposition~\cite{staehle_2025_designspace} inspired the conceptual architecture of \appName.

\subsection{Mixed-Initiative Studies and Study Environments}

When conducting studies for generic VA, there exist approaches that allow for easier interactive data analysis~\cite{cutler_2020_trrack,blascheck_2016_VA2}. VA$^2$ by \citet{blascheck_2016_VA2} focuses on identifying common study participants' interaction strategies after conducting a human subject study with a VA interface. To further equip VA systems with interaction tracking to capture user intent and provide provenance logging, \citet{cutler_2020_trrack} provide the software library Trrack. 
ReVISit~\cite{ding_2023_revisit,cutler_2026_revisit2} provides a study environment that allows for defining web-based visualization studies through a declarative format. The platform has been used for testing, e.g., users' visualization literacy~\cite{nobre_2021_revisitUndertheHood}. Further, researchers can run the defined studies in different settings (lab or crowdsourced study) while tracking user interactions using Trrack~\cite{cutler_2020_trrack} and then allowing further analysis by replaying user sessions. The tool also enables researchers to share existing studies for transparency and replication.
Another web-based framework for conducting behavioral experiments is jsPsych, introduced by \citet{leeuw_2023_jspsych} and currently used by the HCI community~\cite{cumbal_2025_visualizingconfidence,masai_2021_onlineStudy}. It allows researchers to define experiments in self-contained modules that can be reused in other experiments later. Additional community plugins allow capturing participants' interaction, and e.g., support eye tracking.   
ReVISit's setup and jsPsych are primarily tailored for single-participant sessions with isolated visual stimuli and thus do not, in principle, meet the requirements and infrastructure for evaluating and tracking fully interactive mixed-initiative VA situations and systems. Nevertheless, these approaches inspired the study setup configuration (using a modular structure and declarative format) and the initial session replay feature of \appName's Study Environment. 
\section{Design Goals and Requirements}
\label{sec:designGoals}
We derived 10 design goals based on the analysis of prior mixed-initiative VA applications~\cite{zhou_2025_hepha,huang_2026_narrativeScaffolding,messina_2024_guidedVA,sevastjanova_2021_questioncomb}, studies~\cite{narechania_2025_guidancesource,setlur_2022_converseAnalyticalChatbot,Ha_2024_GuidedByAI}, design spaces~\cite{staehle_2025_designspace,Monadjemi_2023_agentbasedFramework,holter_2024_deconstructing}, other VA and HCI study frameworks~\cite{cutler_2026_revisit2,leeuw_2023_jspsych}, and our own assessment of the capabilities we aim to support in \appName. Our overarching goal was to provide a framework for web-based, mixed-initiative VA applications that supports rigorous scientific research, study sharing, and reproducibility. We structure the design goals into goals for \coloredul{colorGeneral}{\textit{General VA Applications and Study Environments}} and \textit{Mixed-Initiative Multi-Agent Frameworks}, which we subdivide into \coloredul{colorInfrastructure}{Infrastructure} and \coloredul{colorStudyEnvironment}{Study Environment} related design goals.

\newpage
 \dimension{\defdesigngoal{G1}{colorGeneral} -- Provenance, Logging and Transparency by Construction}{colorGeneral}
{Every mixed-initiative VA application requires logging of state updates, agent interactions (provenance tracking), and communication to provide transparency. Consequently, the Infrastructure must automatically capture all state updates, agent communications, and interactions with the VA environment to establish complete provenance. These logs should be detailed enough to identify misconfigurations or errors, and to reconstruct a full session~\cite{narechania_2025_guidancesource,cutler_2026_revisit2}. The Study Environment should directly map the logs to the study session and to additional session recordings across all agents, to allow for replay and post-session analysis.}

  \dimension{\defdesigngoal{G2}{colorGeneral} -- Flexible Study Design}{colorGeneral}
    {The technical burden of setting up a study upon an implemented VA environment should be as minimal as possible. Researchers should be provided with an interface for uploading and deploying study configurations. To support studies that combine quantitative and qualitative methodologies, study configurations should be flexible enough to include questionnaires in addition to VA system-related tasks~\cite{leeuw_2023_jspsych,cutler_2026_revisit2}. The configuration should support task definition, including embedding VA interfaces built on the Infrastructure, randomizing the task sequence, implementing attention checks, and handling invalid operations.}

    \dimension{\defdesigngoal{G3}{colorGeneral} -- Sharing of Studies for  Replication of Studies and More Scientific Rigor}{colorGeneral}
{The scientific community is embracing a culture of openness, actively advancing replicable research and shared study designs~\cite{hornbaek_2014_isOneEnough,iarygina_2026_reproducability}. Researchers should be able to use the Study Environment to replicate or extend other researchers' work and share study session analysis results with their stakeholders.}

   \dimension{\defdesigngoal{G4}{colorGeneral} -- Interface for Post-Study Analysis}{colorGeneral}
{Once a study session is completed and data is collected, researchers should be able to see study results, replay the session, explore and analyze the collected data by being provided with an interface that interlinks the data and allows accessing the data for further analysis~\cite{cutler_2026_revisit2}. }  
\newline
\newline
\newline
\noindent Beyond general requirements for study environments, we derived specific goals to enable effective human and agent collaboration, study agent behavior, and advance mixed-initiative systems:   
    
    \dimension{\defdesigngoal{G5}{colorInfrastructure} -- Decoupling Agents from the VA Environment}{colorInfrastructure}
{Following \citet{staehle_2025_designspace}, the Infrastructure should be modular and decouple agents from the VA environment, which would allow researchers to add or replace agents without updating the environment itself. Further, a layer in the Infrastructure should handle environment-state access and manipulation, as well as inter-agent communication, to enable mixed-initiative collaboration~\cite{Horvitz_1999_mixedInitiative}. All modules should support logging. }

    \dimension{\defdesigngoal{G6}{colorInfrastructure} -- Standardized Access for Humans and Software Agents}{colorInfrastructure}
{Given the modular structure, the Infrastructure should provide a structured and standardized approach for adding and implementing agents. 
This should provide an interface and an infrastructure layer for human and software agents to access the VA environment uniformly~\cite{staehle_2025_designspace}.
The infrastructure should allow agents to observe states and perform actions on the VA environment. In addition, if required, a communication module should allow agents to share information directly among each other.}

\newpage   
    \dimension{\defdesigngoal{G7}{colorInfrastructure} -- Centralized and Declarative Configuration of Roles and Permissions}{colorInfrastructure}
{To define and steer agent behavior and permissions, i.e., which parts of the VA environment they can observe, and act on, which actions they can execute, and with which agents they can communicate, it should be possible to configure these permissions and roles through declarative formats to make permission rules easier to audit and update without changing core application code (similar to contributions in Lotse~\cite{sperrle_2023_Lotse}, ReVISit~\cite{ding_2023_revisit,cutler_2026_revisit2}, Vega-lite~\cite{satyanarayan_2017_vegaLite}, and Mosaic~\cite{heer_2024_mosaic}). To support dynamic environments~\cite{staehle_2025_designspace}, the Infrastructure should allow configuration and permission updates at runtime.}

    \dimension{\defdesigngoal{G8}{colorInfrastructure} -- State Synchronization Across Agents and Environment}{colorInfrastructure}
    {As VA application comprises the visual interface, the underlying data, machine learning models, and the users -- the agents -- the state among all components of the application needs to be synchronized at runtime~\cite{staehle_2025_designspace}. I.e., if one agent performs an action that updates the VA environment state, the new state needs to be propagated for updating the underlying data, the visual interface, and possibly the machine learning model. It must be ensured that other agents retrieve the updated state when they observe the VA environment. Further, agents should be able to follow other agents by retrieving their interaction trajectories to react to their actions.}

    \dimension{\defdesigngoal{G9}{colorStudyEnvironment} -- Allow Complex Collaborative Study Setups}{colorStudyEnvironment}
{To study agentic mixed-initiative VA systems, researchers apply a diverse set of study scenarios~\cite{wobbrock_2016_researchcontributionHCI}, such as general user studies for collaboration and interaction logging~\cite{zhao_2026_proactiveVA}, controlled user studies with multiple conditions to measure the impact of different agent types or behaviors~\cite{narechania_2025_guidancesource}, or different VA interfaces on the collaborations~\cite{kuhlman_2019_evalPreference}, or Wizard-of-Oz (WoZ) scenarios~\cite{sperrle_2025_WoZ}. Further, common methods for data collection in HCI and VA are lab-based or crowdsourced studies. The Study Environment should support flexible collaborative study-scenario design and settings that allow scaling the number of collaborators in a study effortlessly.}

    \dimension{\defdesigngoal{G10}{colorStudyEnvironment} -- Advanced Data Collection }{colorStudyEnvironment}
    {Besides participants' interaction and behavior data, researchers are more and more interested in collecting additional data, such as biometric sensor data (heart rate, etc.)~\cite{ellemose_2026_frustrometer,panwar_2018_negativeEmotions,krisam_2026_mindtide}. The Study Environment should support researchers in collecting additional sensor data during a study session, e.g., to later analyze participants' stress and cognitive load.}

\section{The \appName Architecture}
\appName builds on the frameworks of agent-centric views~\cite{Monadjemi_2023_agentbasedFramework,holter_2024_deconstructing}: VA systems consist of a VA environment in which human and software agents operate (\refdesigngoal{G5}{colorInfrastructure}). This allows for deconstructing mixed-initiative VA systems into subparts that build the infrastructure and agents, enabling a modular approach to standardize the design of mixed-initiative systems and facilitating easier system comparison, configuration, and evaluation. In the following, we describe the architecture and components of \appName provided in \Cref{fig:teaser}, explicitly mapping each component to its corresponding design goal \refdesigngoal{Gx}{colorGeneral} (defined in \Cref{sec:designGoals}).

\begin{figure}[h]
  \centering
 \includegraphics[width=\linewidth]{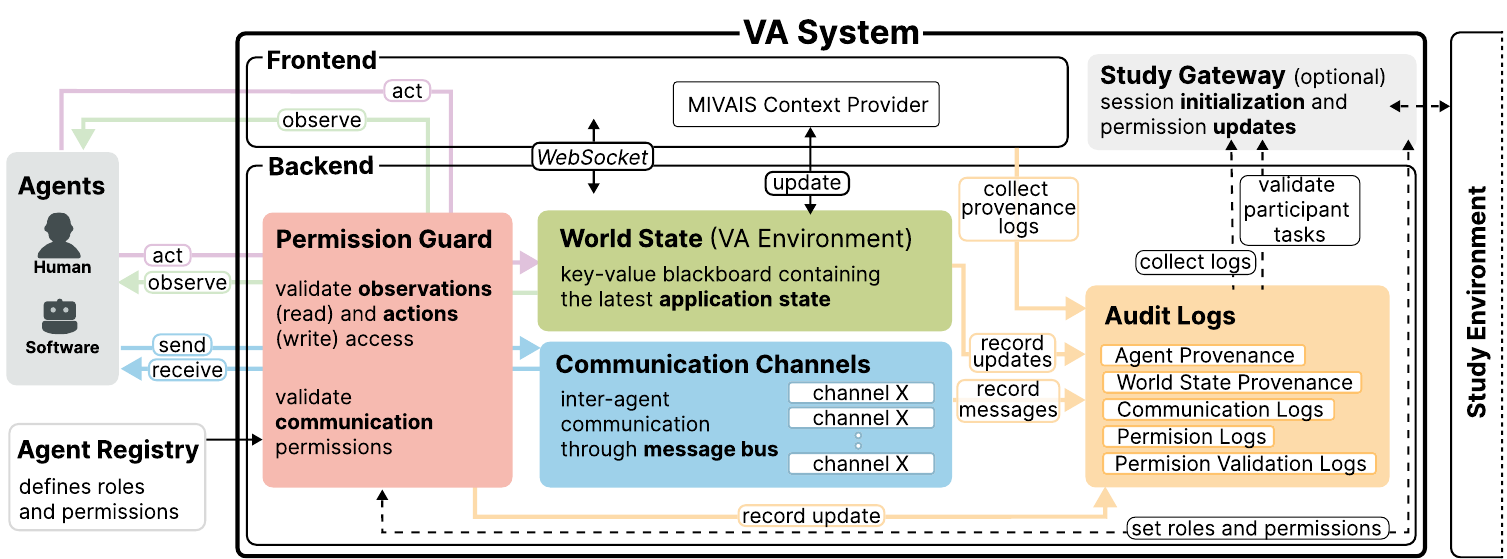}
  \vspace{-2em}
  \caption{Architecture and data flow of a VA system built on the \appName Infrastructure. Human and software agents interact with the application's World State and each other, strictly mediated by a central Permission Guard. The backend handles state management and comprehensive audit logging, while an optional Study Gateway facilitates seamless integration with the broader Study Environment.%
  }
  \label{fig:Infrastructure}
  \Description{In a system built on the \appName Infrastructure, agents observe and act on the World State through the frontend or directly through the backend. Agents can send and receive messages through Communication Channels. The Permission Guard validates interactions with the World State and inter-agent communication based on the permission defined in the Agent Registry. The visual interface (frontend) and the backend are connected via a WebSocket communication that updates the state across the \appName Context Provider (frontend) and World State in the backend. At runtime, Audit Logs are captured by collecting provenance and communication data from the agents and recording World State and permission updates. If the VA system is used with the Study Environment, the Study Gateway connects with the Study Environment and can be used to initiate a study session by setting agent roles, permissions, and initial World State. Further is collects the logs and can automatically validate participant tasks based on interactions or state updates.}
   \vspace{-1em}
\end{figure}
\subsection{The Mixed-Initiative Infrastructure Model}
To enable the researcher to build mixed-initiative VA systems with reduced effort, the \appName Infrastructure (\Cref{fig:Infrastructure}) comprises the six modular components, explained in the following: VA Environment (World State), Agents, Agent Registry, Permission Guard, VA Visual Interface, and Agent Communication Channels.
\subsubsection[VA Environment and World State]{\protect\defcomponent{World State and VA Environment}{colorComponentWorldState}}
Similar to other works~\cite{Monadjemi_2023_agentbasedFramework, Abel_2023_ContinualReinforcementLearning,staehle_2025_designspace}, the VA environment contains components, such as data management, analytical models, visualizations, knowledge representation, and potentially data collected through evaluation or feedback loops. The environment is built on the World State. The World State represents the current state of the VA environment at any given time (\refdesigngoal{G8}{colorInfrastructure}). Therefore, all data and entities that can be updated through actions on the environment are stored in the World State. E.g., if an agent can select a dataset and specify a chart for further exploration, the dataset and the chart specification are stored in the World State.   
Every agent synchronizes with the World State to consume (observe) and update (act) the latest state. This is crucial for mixed-initiative applications where multiple agents can operate in the environment independently on their own schedule, influencing the World State. Further, a synchronized World State is important, as agents might join or leave the session during runtime (either through dynamic updates to the application configuration or through human agents quitting the session). If an agent joins an ongoing session, only consuming new events is not sufficient without full access to the underlying World State, as it needs the relevant context to evaluate the effects of those events. We operationalize the World State through a central and shared ``blackboard'' where every agent can read and update the state.     
Without any permissions, every agent interacting with the VA environment can subscribe to and update the World State.

Although the World State could also be stored in a decentralized manner, the centralized approach offers multiple advantages such as more straightforward  permission management, data collection, and provenance logging when conducting research studies (\refdesigngoal{G1}{colorGeneral}).  
In addition to tracking the current World State, a provenance log of all manipulations is created.  
\subsubsection[Agents]{\protect\defcomponent{Agents}{colorComponentAgents}}
Based on the definition by \citet{staehle_2025_designspace}, in mixed-initiative (VA) systems, an agent is any entity, human or software, ``capable of perceiving its environment and proactively performing goal-driven actions. An agent must be capable of autonomous decision-making by interpreting its environment and applying internal logic, allowing it to take or cede initiative.'' In line with this definition, in \appName every agent is modeled uniformly. Agents can perceive the full or parts of the VA environment through accessing the World State or listening to its updates (\refdesigngoal{G5}{colorInfrastructure}). Further, any logic within the agents can derive a decision and act on the environment by updating parts of the World State. When implementing an agent, it is important to define which parts of the environment (World State) the agent needs to observe, how it computes and derives its next actions, and which parts of the environment (World State) are affected by these actions (\refdesigngoal{G6}{colorInfrastructure}). 
Agents can communicate with other agents by sending and receiving messages in subscribed channels (\Cref{sec:CommunicationChannels}). 
This allows agents to notify each other, request actions, or share information. Each software agent can be implemented in a separate code file that defines its logic and is activated once registered in the Agent Registry.    
\subsubsection[Agent Communication Channels]{\protect\defcomponent{Agent Communication Channels}{colorComponentCommunication}} \label{sec:CommunicationChannels}
To allow agent-to-agent interactions, \appName provides Communication Channels. As messages across agents do not update the World State, by design, these signals do not pass through the environment (\refdesigngoal{G6}{colorInfrastructure}). Instead, agents can subscribe to and publish to communication channels. This allows us to design any kind of agent communication, e.g., distributed, centralized, or hierarchical networks. Further, these networks can be dynamically updated at runtime as agents can dynamically (un)subscribe to channels. The body of messages is text-based, allowing human and software agents to communicate in natural language. At the same time, this also supports structured language for messages only sent between software agents. Because the central infrastructure of \appName has access to all channels, it is possible to log all messages exchanged between agents (\refdesigngoal{G1}{colorGeneral}).    

\subsubsection[Agent Registry and Permission Guard]{\protect\defcomponent{Agent Registry and Permission Guard}{colorComponentPermissionGuard}}
The Agent Registry uses a domain-specific language (DSL) to define which agents are present in the system, specifying their identifiers and roles (Appendix \ref{appendix:dsl_agent_definition}). Additionally, it defines the permissions, i.e., which agent can access and update which parts of the World State (\refdesigngoal{G7}{colorInfrastructure}). Agents and Permissions can be dynamically added or removed from the registry. This allows researchers and developers to manipulate the behavior and scenarios of their applications at runtime using scheduled conditions or updates on demand.
The \textit{Permission Guard} centralizes the validation of permissions defined in the Agent Registry. Thus, before agents can perceive or update parts of the World State, the operation is validated. If an agent tries to access or update a part of the VA environment for which it has no permission, the permission guard denies the request and informs the agent. If the agent has permission, the Permission Guard lets the agent's observation or state update pass. Further, the Permission Guard ensures that agents can send and receive messages only through the communication channels they are authorized to use.  
This centralized access control enables auditing of all access, update, and communication attempts by agents (\refdesigngoal{G1}{colorGeneral}). %
\subsubsection[Visual Interface for Agents Relying on the UI]{\protect\defcomponent{Visual Interface for Agents Relying on the UI}{colorComponentInterface}}
Human visual perception is particularly effective at recognizing complex patterns that can be difficult for algorithms to detect. VA environments therefore use graphical interfaces to represent their World State. To interact effectively with these environments, both human and software agents (e.g.,~\cite{staehle_2026_vacp}) require direct visual access to this representation. In this work, we focus on web-based interfaces that allow us to use technologies such as WebSockets for state synchronization between the application's backend and frontend. When an agent executes an interaction to act upon the VA environment in the visual interface, this interaction is first validated by the Permission Guard, and if valid, updates the World State. When another agent updates the World State, the visual interface updates the visual representation to present the updated state (\refdesigngoal{G8}{colorInfrastructure}). As UI agents can perform interactions in the visual interface, such as mouse interactions that contain rich information~\cite{nguyen_2016_sensemap,holter_2026_intent2interaction}, these interactions can be recorded, displayed in the interface of another agent, and passed to the central audit log of \appName (\refdesigngoal{G1}{colorGeneral}).  

\subsubsection{Summary} Grounded in the design goals, these components simplify mixed-initiative VA system development, allowing researchers and developers to focus on agent logic while abstracting away state synchronization, communication, and permission management. We show its applicability in \Cref{sec:replicationWork}. 

\subsection{The Study Environment}
Once researchers have implemented new systems or approaches, the main goal is to evaluate these. Similar to many other research areas in HCI and VA, mixed-initiative approaches are usually evaluated based on a mix of quantitative and qualitative methods. Researchers, therefore, use multiple tools in a queue to set up studies and enable the necessary data collection (e.g., by tracking user interactions and final application states). Furthermore, they calculate metrics on application logs, record the screen of the participant(s), capture think-aloud sessions, and collect questionnaire results, commonly through a separate tool, such as Google  Forms\footnote{\href{https://docs.google.com/forms}{https://docs.google.com/forms}}. This creates significant overhead, while leaving collected data fragmented across multiple sources, complicating data aggregation and alignment for downstream analysis. It becomes even more challenging when testing multiple applications, running multi-user sessions, or conducting WoZ experiments.  
When designing \appName, we address this gap by following best practices from prior work and introducing new features that are especially relevant for evaluating multi-agent mixed-initiative systems.
For each feature, we first outline its similarities to existing study frameworks, followed by the unique properties \appName provides for mixed-initiative setups.      
\subsubsection{Study Setup}
Studying mixed-initiative approaches in VA can have many different facets~\cite{wobbrock_2016_researchcontributionHCI,sperrle_2021_humancentered}. Therefore, before conducting a study, researchers need to specify the experiment. 
Like existing frameworks~\cite{cutler_2026_revisit2}, \appName uses a DSL to define study structures, participants, and tasks (\refdesigngoal{G2}{colorGeneral}), although in \appName, we use \texttt{YAML} instead of \texttt{JSON} to support inline comments for documentation. Researchers can edit study and task files directly in the Study Environment interface or upload them from an IDE. Studies include ID tags, names, optional versioning for identification, and task sequences, explained in \Cref{sec:task-definition}. %
Beyond interaction and response tracking (\refdesigngoal{G1}{colorGeneral}, \Cref{sec:provenance_tracking,sec:tracking_screen}), \appName provides built-in multi-modal collection, including screen recording, audio capture, and automatic transcription (\refdesigngoal{G10}{colorStudyEnvironment}). Finally, researchers can configure an integrated consent form shown before the experiment to address ethics and privacy (\refdesigngoal{G1}{colorGeneral}).

Additionally, \appName provides several unique properties for study configuration.
For instance, it allows researchers to conduct both single- or multi-system evaluations (\refdesigngoal{G9}{colorStudyEnvironment}) simply by referencing the target systems in the DSL during setup. Systems can be hosted internally or externally, requiring a reference name, system URL, and health endpoint. The URL embeds the VA interface into the participant view and enables backend communication with the \appName Study Gateway (\Cref{fig:Infrastructure}), while the health endpoint tracks availability for logging and crash recovery.
To support single- or multi-participant studies, researchers specify the participant mode, total capacity, and specific roles (e.g., user vs. wizard in WoZ studies) (\refdesigngoal{G9}{colorStudyEnvironment}). \appName captures multi-modal data from all collaborating participants simultaneously and aligns it along a shared timeline (\refdesigngoal{G10}{colorStudyEnvironment}).

\subsubsection{Task Definition}\label{sec:task-definition}
\begin{figure}[!t]
  \centering
 \includegraphics[width=\linewidth]{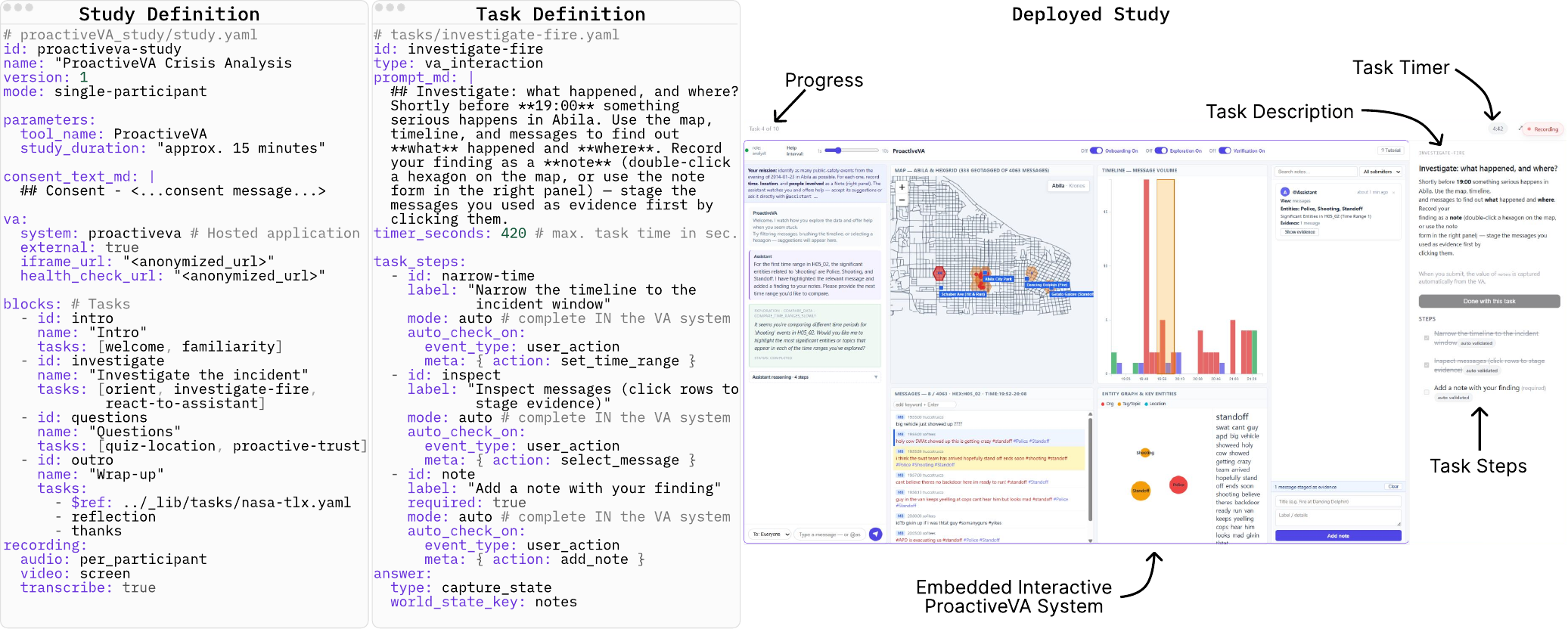}
  \vspace{-2em}
  \caption{In \appName, a study and its tasks are defined in DSL \texttt{YAML} files. This example shows an excerpt from the Study Definition and one of the Task Definitions for a study on the reimplemented ProactiveVA system (with additional configuration details elided). The right part shows the deployed study and task, as well as the resulting study interface provided to participants.}
 \label{fig:DSL-definition}
  \Description{In \appName, a study and its tasks are defined in DSL \texttt{YAML} files. This example shows an excerpt from the Study Definition and one of the Task Definitions for a study on the reimplemented ProactiveVA system. The right part shows the deployed study and task, as well as the resulting study interface provided to participants.}
   \vspace{-1.5em}
\end{figure}

Tasks define the study's subparts and can vary in structure. %
For modularity and reuse, tasks are stored in stand-alone \texttt{YAML} files identified by a task ID and a relative path (optional), which allows reusing them multiple times (e.g., providing participants with the NASA-TLX~\cite{hart_1988nasaTlx} or System Usability Scale~\cite{Brooke_1996_SUS}). 
In \appName, task sequences are grouped into blocks and subgroups, which are processed sequentially or shuffled to minimize carry-over effects~\cite{lazar_2017_researchMethods,brooks_2012_counterbalancing} (\refdesigngoal{G2}{colorGeneral}).
Similar to reVISit and jsPsych, \appName allows researchers to display the stimulus, i.e., a target mixed-initiative VA application, alongside task descriptions (see \Cref{fig:DSL-definition}), interactive forms (e.g., subtask lists, finding reports, or Likert scales), questionnaires, or attention tests (\refdesigngoal{G2}{colorGeneral}). If required, tasks can also be timed to enforce completion deadlines.
Additionally, we allow storing tasks in a central library for reuse across multiple studies, eliminating redundant configuration (\refdesigngoal{G3}{colorGeneral}). 
The DSL grammar for a task definition is provided in Appendix \ref{appendix:dsl_task_config}.

To address mixed-initiative setups, \appName uniquely enables researchers to either reconfigure agent permissions and initialize the VA system's World State for individual tasks (e.g., disabling an agent from offering proactive visualization recommendations during an initial exploratory task to establish a baseline for unassisted user discovery) or maintain a continuous system session across sequential tasks. Additionally, tasks in \appName can include subtasks that must be completed within the linked mixed-initiative VA application (\refdesigngoal{G9}{colorStudyEnvironment}), which are validated automatically by tracking the application's event logs received through the Study Gateway and monitoring for a certain event or accumulation of events (\Cref{fig:DSL-definition}).

\subsubsection{Provenance Tracking}
\label{sec:provenance_tracking}

As with other study frameworks, \appName records mouse interactions and cursor paths, and it continuously collects provenance trace logs, storing them together with participant responses and telemetry (\refdesigngoal{G1}{colorGeneral}). This setup enables both high-level summaries and detailed, on-demand inspection during post-session analysis.

In addition, the Study Environment's provenance tracking of all agents is built on centralized audit logs created in each application that uses the \appName Infrastructure. 
Once a study is deployed and participants begin a task using a \appName-based VA application, the Study Environment connects directly to the application's Study Gateway via WebSocket and collects the data. In addition, it allows researchers to capture custom agent-interaction events. Beyond agent provenance, the system tracks World State changes and inter-agent communication. Any additional data recorded by researchers, like biometric sensor data, is automatically aligned with these provenance traces (\refdesigngoal{G10}{colorStudyEnvironment}).

\subsubsection{Tracking of Interface and Audio}
\label{sec:tracking_screen}

HCI and VA research frequently incorporates screen and audio recordings. Capturing these media streams allows researchers to record the exact visual stimuli perceived by participants alongside think-aloud protocols during task execution. When configuring a study with \appName, researchers can enable screen and audio recording. As our framework is web-based, this feature uses the built-in browser's APIs to record the screen and audio. The batched recording chunks are send from the participant's client to the host of the Study Environment. If participants decline the recording, the study can be terminated immediately or continued without recoding the data. Similar to reVISit, \appName provides automatic transcription of recorded audio. Transcripts are aligned with the other tracking data such that timestamps match and can be used for coding~\cite{Braun_2006_thematicAnalysis} or further analysis of participants' thoughts~\cite{erricsson_1993_protocolAnalysis}.

To address mixed-initiative needs, \appName can capture the interface and audio of all participants in the same session simultaneously and align them on a timeline, allowing switching point of view at any time during post-analysis (\refdesigngoal{G10}{colorStudyEnvironment}).

\subsubsection{Tracking of Additional Sensor Data}
To afford maximum flexibility in data collection (\refdesigngoal{G10}{colorStudyEnvironment}) in mixed-initiative setups, \appName offers a REST API endpoint capable of ingesting external data streams into active study sessions. Researchers can utilize this mechanism to capture additional metrics, such as biometric sensor data or eye-tracking logs gathered from secondary devices, and consolidate them for offline analysis. The endpoint can handle single and batched data. This allows for reducing data traffic while still enabling high-frequency data collection. Additional data tracking can be enabled in the study configuration. If activated, a unique token is provided at the beginning of the study session that researchers add to their sending devices. The token takes care of authentication and study session mapping. Data can be added both during an active session and after the session is completed to add data from offline sensors. The data is aligned to the other data through the ISO-8601 wall-clock timestamp.

\subsubsection{Deploying and Executing Studies}
Addressing our design goal \refdesigngoal{G2}{colorGeneral}, the Study Environment provides researchers with necessary infrastructure to deploy their studies. 
Once a study is designed, configured, and the configuration files are uploaded to the \appName Study Environment, researchers register a study for deployment in the admin visual interface. All related study and task configurations are copied to a read-only directory to preserve the underlying files from being updated while the study is deployed. \appName covers hosting the study, providing each participant with a unique access code. Participants access the study by visiting the base URL of \appName and entering their code (this forwards the participant to the linked study).
Studies can be conducted through \textit{lab studies} or \textit{crowdsourcing} (\refdesigngoal{G9}{colorStudyEnvironment}). For \textit{lab studies}, researchers can manually generate and share the access codes with the participants. Participants then access the study via a web browser on a computer in the lab setup. 
To run \textit{crowd-sourced studies}, \appName currently provides a connector for Prolific\footnote{\href{https://www.prolific.com/}{www.prolific.com/}} that lets researchers source participants through the Prolific Platform by lodging a study-specific URL. The connector handles assigning participants to the correct study session, storing their participant IDs and other metadata, and, at the end, forwarding them to the Prolific Platform. As there are multiple more providers for crowd-sourced studies, we plan to provide additional connectors soon.
An experiment can be terminated either when a predefined number of participants is reached, which prevents other participants from signing up, or when researchers manually terminate the study by archiving it. 
For investigation and analysis of a single completed study session, it is not required to terminate and archive the whole study. 

As an additional feature, \appName validates the syntax of the study and task configuration files and, if correct, deploys the study immediately. 
If a study contains multiple participants who should collaborate in the same session, we also provide mini-cohorts (\refdesigngoal{G9}{colorStudyEnvironment}): access codes directly forward participants into a shared session. \appName isolates each session and propagates its context to the involved VA systems. 
During tasks, the \appName Study Environment connects directly to the active VA session to collect and store provenance and event data for post-session analysis. Researchers can also passively monitor active sessions in real time via the \textit{Live} view.

\subsubsection{Post-Study Session Replay and Evaluation}
Following study completion, \appName's Study Environment provides several features to assist researchers with data analysis and evaluation.

\begin{figure}[t]
  \centering
 \includegraphics[width=\linewidth]{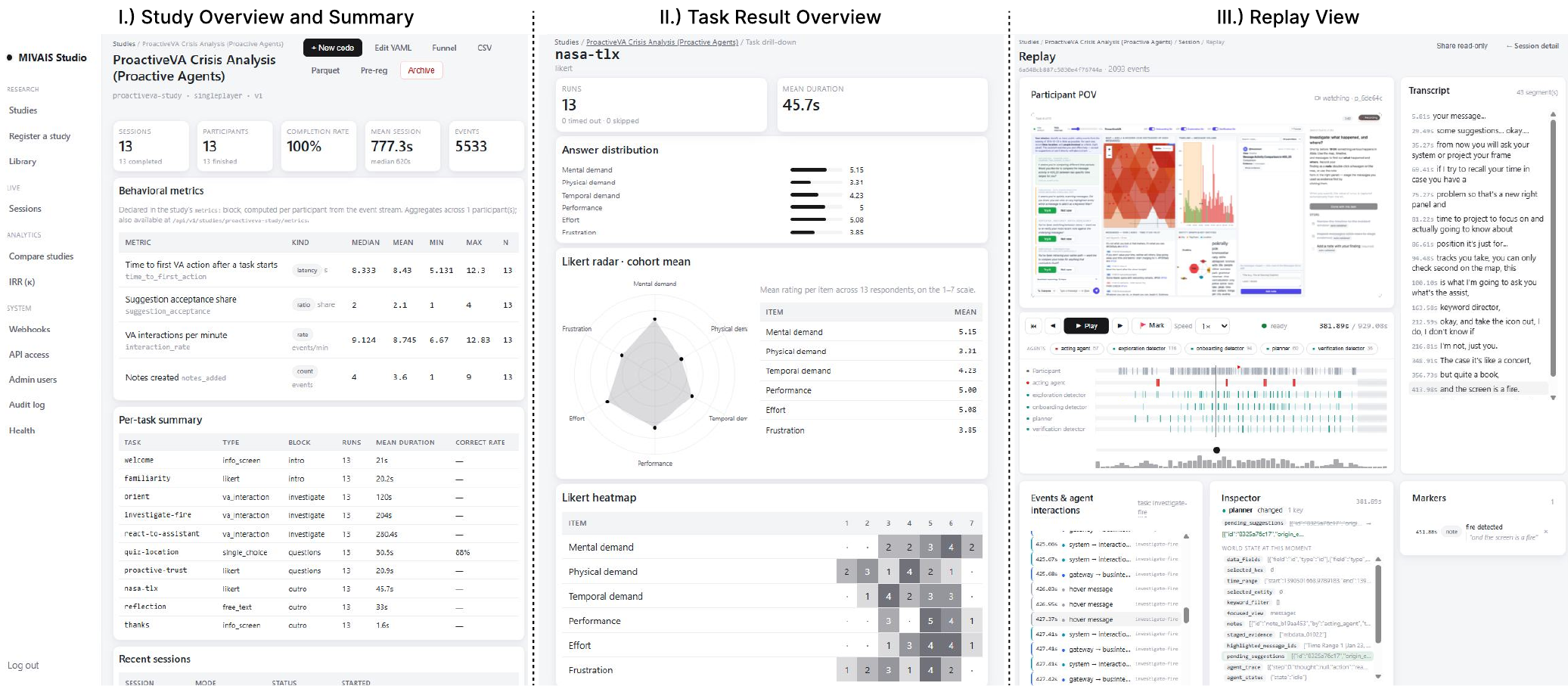}
  \vspace{-2em}
  \caption{The visual interface of \appName for post-study analysis. The dashboard provides researchers with three primary analytical views: (I.) a Study Overview summarizing aggregate metrics and session progress, (II.) a Task Result Overview for evaluating specific task outcomes (e.g., questionnaires), and (III.) a comprehensive Replay View that synchronizes screen recordings, interaction timelines, and transcripts for deep inspection of individual sessions.%
  }
 \label{fig:MIVAIS_interface}
  \Description{The visual interface of MIVAIS for post-study analysis. The dashboard provides researchers with three primary analytical views: (I) a Study Overview summarizing aggregate metrics and session progress, (II) a Task Result Overview for evaluating specific task outcomes (e.g., questionnaires), and (III) a comprehensive Replay View that synchronizes screen recordings, interaction timelines, and transcripts for deep inspection of individual sessions.}
   \vspace{-1.5em}
\end{figure}

\paragraph{Session Overview and Participants' Responses} 
Consistent with existing study platforms, selecting a study with completed sessions displays high-level statistics in the \textit{Study Overview} and \textit{Summary View} (\Cref{fig:MIVAIS_interface}), including session counts, task and subgroup summaries, completion times, and aggregated participant responses (\refdesigngoal{G4}{colorGeneral}). Selecting an individual task opens the \textit{Single Task Result Overview} (\Cref{fig:MIVAIS_interface}), showing task-specific participant responses or performance metrics.

\paragraph{Interaction Timeline and VA Environment State Analysis}
To support multi-agent collaboration analysis, the timeline visualizes all human and software agent interactions, communications (\refdesigngoal{G8}{colorInfrastructure}), and recorded sensor data during VA system tasks (\refdesigngoal{G10}{colorStudyEnvironment}). Researchers can define custom interaction or event sequences to be highlighted on the timeline (\refdesigngoal{G3}{colorGeneral}), revealing individual agent behaviors or multi-agent interaction patterns.

\paragraph{Session Replay} 
Additionally, researchers can review individual study sessions using the \textit{Replay} feature (\Cref{fig:MIVAIS_interface}, III.). To analyze how agent actions transform the World State, the \textit{Replay View} displays the state at any timestep, highlighting updated values and allowing researchers to click variables to compare previous and current values. In multi-participant sessions, researchers can toggle between participant viewpoints, with the active view indicated by a camera icon next to the participant's ID.

\paragraph{Think-Aloud Analysis and Annotation}
When audio recording is enabled, \appName automatically transcribes input audio with timestamp annotations using Whisper small~\cite{radford_2022_whisper} (\refdesigngoal{G4}{colorGeneral}). As \citet{cutler_2025_thinkaloud} note, transcripts in think-aloud studies aid in insight elicitation, detecting participant surprise during agent interactions, and usability testing. Researchers can highlight and annotate specific transcript sections, which are then reflected on the timeline visualization. Similar to \citet{cutler_2026_revisit2}, this captures user thought processes during tasks rather than post-interaction.

\paragraph{Further Analysis}
To enable deeper analysis of the collected data, \appName offers the option to export the data as \texttt{JSON} lists, which can then be imported into any analysis tool or notebook the researcher prefers.

\subsection{Library Implementation and API}
The Infrastructure of \appName is implemented as a Python library that provides the base infrastructure for the described architecture and a WebSocket interface for the frontend. We further provide a TypeScript library that leverages backend communication in the frontend of the application, leaving the choice of frontend framework to the developer of the mixed-initiative VA system. 
The Study Environment is based on a FastAPI backend, providing a RESTful API and storing data in a MongoDB. The visual interface is implemented in React TypeScript. For deployment, we provide Docker containers that can be hosted or run locally. The source code is provided in the supplemental material.
   
\section{Technical Evaluation}
A typical \appName workflow begins with conceptualizing the system by outlining the core data model, machine learning components, and VA interface. Next, researchers implement this draft in the backend by defining the central World State, agent configurations, permissions, and communication channels. After coding the agents' logic and integrating the frontend via the WebSocket API, the system is complete. Finally, researchers use the Study Environment to configure, deploy, and analyze their study before sharing their system and results with the community for replication. A more detailed description is provided in Appendix \ref{appendix:usageSecenario}.

We evaluate \appName's applicability by adapting three prior mixed-initiative VA interfaces to our Infrastructure and assess our design using the Technical Dimensions of Programming Systems (TDPS)~\cite{Jakubovic_2023_TDPS}.

\subsection{Replicating Prior Work}
\label{sec:replicationWork}
We utilize \appName to replicate three mixed-initiative prior works in 
informing data-driven decisions (Podium~\cite{wall_2018_podium}), data exploration and targeted question answering (Voyager~2~\cite{wongsuphasawat_2017_voyager2}), and context-aware proactive assistance in VA (ProactiveVA~\cite{zhao_2026_proactiveVA}). The source code is provided in supplemental material.

\subsubsection{Podium}
\citet{wall_2018_podium} developed Podium, a mixed-initiative VA application that helps users identify attributes relevant to ranking multivariate data. Users drag table rows to express their preferred ranking, and a Ranking-SVM–based agent infers an attribute-weighting model that best fits these preferences. 

We reimplement Podium’s interface and functionality so human users can rank cars by their attributes, given a dataset of automobile statistics~\cite{quinlan_1993_cardataset}. Our implementation maintains a \refcomponent{World State}{colorComponentWorldState} that stores the initial order of data points and attribute weights. It also includes two \refcomponent{agents}{colorComponentAgents} (one human, one software), both with \refcomponent{permissions}{colorComponentPermissionGuard} to update the ranking in the World State, while only the human can add priorities to specific attribute weights and only the software agent can change the attribute weights in the World State. The human agent can request the software agent to update attribute weights or rank all data points using the current weights by sending a message through the \refcomponent{communication channel}{colorComponentCommunication} (triggered by a button click). In the \refcomponent{VA interface}{colorComponentInterface}, we bind the World State ranking to the main table and map attribute weights and priorities to the control panel visualization.

\subsubsection{Voyager~2} \label{sec:reimplementationVoyager}
By combining manual and automated chart specification, Voyager~2~\cite{wongsuphasawat_2017_voyager2} enables analysts explore data and drill down to specific questions. Users create charts to find insights, while a software agent simultaneously suggests additional visualizations relevant to the current chart.

In our \appName-based reimplementation of Voyager~2, the \refcomponent{user interface}{colorComponentInterface} provides the same views and functions as the original. The views synchronize relevant data with the \refcomponent{World State}{colorComponentWorldState}, which stores uploaded datasets, current chart specifications, alternative encodings, related summaries, wildcard results, field suggestions, and bookmarked charts. We have two \refcomponent{agents}{colorComponentAgents} (one human, one software). \refcomponent{Permissions}{colorComponentPermissionGuard} match the original: the human agent can manually select attributes and wildcards to define charts and bookmark specifications. The software agent generates insights and recommends alternative encodings and field suggestions based on the human’s current specification. Recommendations update the World State and thus appear in the user interface as selectable, bookmarkable options, whereas generated insights are sent directly to the human analyst via the \refcomponent{communication channel}{colorComponentCommunication}.

To assess the feasibility of multiple software \refcomponent{agents}{colorComponentAgents} in this environment, we implemented a variant that splits insight and recommendation generation between two agents (recommendation and insight) with different \refcomponent{permissions}{colorComponentPermissionGuard}. The recommendation agent suggests related views by observing the \refcomponent{World State}{colorComponentWorldState} (focus view and chart specification) and updating alternative encodings, field suggestions, and related summaries. The insight agent operates only on the dataset, updating focus encodings and \refcomponent{communicating}{colorComponentCommunication} identified insights to the human user, including possible actions. If the human agent selects an action, it is sent to the insight agent, which then instructs the recommendation agent to apply it in subsequent suggested chart generations.

\subsubsection{ProactiveVA}
\citet{zhao_2026_proactiveVA} introduce ProactiveVA, a framework using an LLM-powered software agent that proactively offers context-aware assistance by monitoring human interactions. We reimplement their social event analysis system using the \appName infrastructure.

To provide a similar \refcomponent{interface}{colorComponentInterface} and identical functionality to the human user as the original system, we implemented ProactiveVA with a \refcomponent{World State}{colorComponentWorldState} that stores the static dataset, selected entities and hexagons on the map, time range, filter keywords, current focused view, notes, pending suggestions, and staged evidence. In the original design, a single software agent detects when help is needed, infers user intent, provides suggestions, and, once confirmed, reasons over and acts on the data. This agent perceives the environment (\textit{observation}), reasons for decision-making (\textit{logic}), and acts (\textit{actions}) on the environment.

Although \appName could replicate this monolithic agent, we intentionally adopted a multi-agent architecture as an illustrative case study: five \refcomponent{agents}{colorComponentAgents} execute the subtasks in~\cite{zhao_2026_proactiveVA} and \refcomponent{communicate}{colorComponentCommunication} via shared channels. We do not claim this decomposition is superior for ProactiveVA; instead, we split the logic to explicitly demonstrate \appName's agent-to-agent communication and decentralized coordination -- core framework features a monolithic design would not validate.
Our implementation uses three intent-perception agents (help-needed detector, user-analysis checker, possible-solution detector), each \refcomponent{provided access}{colorComponentPermissionGuard} to different parts of the \refcomponent{World State}{colorComponentWorldState} to detect when the user needs assistance. Two additional agents (planner and actor) generate suggestions and intents, reason over the data, and adjust data selections. As the human \refcomponent{user}{colorComponentAgents} explores the data, the perception agents analyze the user's behavior. When they detect a need for assistance, they notify the planner via the assistance-required communication channel. Receiving the message, the planner generates suggestions and sends them to the user via the pending-suggestion channel. After the human confirms a suggestion, the actor executes it. The user can also directly message the actor via chat to request help. Unlike the original ProactiveVA, we use Gemini 2.5 flash~\cite{gemini2023geminiFamily} for LLM inference.     

\subsubsection{Summary and Lessons Learned}
The original systems we replicated differ in nature and functionality, yet all can be implemented on the same modular Infrastructure. The replication process shaped the final synchronization mechanism and permission guard: we moved from broadcasting all state updates to all agents without restriction to a guarded, selective update scheme. It also refined the structure of agent configuration files, revealing additional useful features, such as hyperparameter injection to steer agent behavior. Although first decomposing an existing application into the core components needed for the \appName Infrastructure is cognitively demanding, we could quickly adapt agents in an existing system (\Cref{sec:reimplementationVoyager}) by leaving most of the code unchanged and focusing on agent implementation and reconfiguration.

\subsection{Technical Dimensions of Programming Systems (TDPS)}
Similar to \citet{cutler_2026_revisit2}, we self-assess \appName via close reading utilizing the ``Technical Dimensions of Programming Systems'' framework~\cite{Jakubovic_2023_TDPS} and the \textit{cluster descriptions} by \citet{mcnutt_2025}. 
To provide analytical clarity, we decouple our assessment into the \textbf{Infrastructure} and \textbf{Study Environment}. The first author assessed \appName for each of the 7 dimensions, and two other authors reviewed it.

\subsubsection{Interaction} \textit{Which loops in the system are overlapping and how far apart are the corresponding gulfs of evaluation?}

\textbf{Infrastructure:}
\appName narrows the traditionally wide feedback loops inherent in mixed-initiative VA development. The modular architecture enables software agents to be implemented and tested sequentially, independent of the VA interface and the study setup.

\textbf{Study Environment:}
\appName mitigates the wide, iterative feedback loops in study design by enabling researchers to test isolated subtasks without executing a full study walk-through. The architecture manages session isolation and data storage without requiring the VA application to be explicitly deployed within the Study Environment.

\subsubsection{Notation} \textit{What notations are present and how do they interrelate?}
We utilize a \texttt{YAML}-based DSL to define agent and study configurations.
The \appName DSL is based on a common abstraction to avoid the need to maintain a full programming language, which might result in a steep learning curve and increased difficulty in generating configurations.
We explicitly decided on \texttt{YAML} as, unlike \texttt{JSON}, it allows for adding comments within the files to share additional information with other stakeholders or researchers.

\textbf{Infrastructure:}
Developers configure modular agent permissions using highly composable \texttt{YAML} files, maintaining the freedom to build VA interfaces using their preferred frameworks.

\textbf{Study Environment:}
The DSL enables researchers to sequence tasks and dynamically overwrite agent parameters for ablations without recompiling the underlying application. While \texttt{YAML} is structurally prone to indentation errors, \appName actively mitigates this risk through strict pre-deployment schema validation.

\subsubsection{Conceptual Structure} \textit{What is the shape of the notations at play and how do they relate to user goals?}

\textbf{Infrastructure:}
\appName enforces a modular conceptual paradigm that programmatically unifies human and software agents. To manage the inherent temporal mismatches and race conditions when software agents and humans interact simultaneously, all updates are routed through a centralized World State and validated by a Permission Guard.

\textbf{Study Environment:}
Leveraging declarative configurations for permissions, tasks, and studies aligns with previous works~\cite{sperrle_2023_Lotse,ding_2023_revisit}. Further, it centralizes study execution and multimodal data tracking for a diverse set of study setups common in HCI and VA, such as system testing, human-software collaboration, multi-agent interaction, Wizard of Oz studies, etc. Nevertheless, relying on the infrastructure and specification of the study through a declarative format might not satisfy all users.

\subsubsection{Customizability} \textit{How can programs be modified?}

The entire codebase will be open-source, but we explicitly avoid relying on ``forking'' as a primary extension mechanism to maintain structural standardization. Nevertheless, in exceptional cases, if required, users can investigate and change the full source-code.

\textbf{Infrastructure:}
The \appName architecture is designed for modular extension; developers can dynamically add custom agents, update permissions, or define tracking events without modifying the core infrastructure.

\textbf{Study Environment:}
Researchers can easily share, replicate, and modify studies by customizing task orders in the configuration files. Furthermore, an API supports integrating supplementary data streams, such as biometric sensors, directly into the session log.

\subsubsection{Complexity}\label{sec:TDPScomplexity} \textit{How is complexity dealt with through design and automation?}
\appName explicitly separates application logic from study execution to reduce cognitive overload.

\textbf{Infrastructure:}
The strict modularity isolates agent definition, state synchronization, and inter-agent communication, allowing developers to focus entirely on agent and environment logic.

\textbf{Study Environment:}
The declarative setup lowers study implementation effort but introduces an integration trade-off: researchers must maintain two separate mental models (application versus study setup), requiring careful alignment when configuring cross-system dependencies.
This duality is inherent to mixed-initiative research: the engineering model of application differs from the experimental model of study tasks and metrics. \appName preserves this separation with a strict architectural boundary, preventing study-specific task logic or logging from contaminating the application code.

\subsubsection{Errors} \textit{What are they and how are they handled?}
Error handling depends on the error that occurs and whether it is related to the Infrastructure or the Study Environment. 

\textbf{Infrastructure:}
The modularity isolates malfunctioning agent behaviors, while the type-safe \appName libraries utilize linters to expose invalid mappings. Built-in audit logs can expose agent misconfigurations, unauthorized update attempts, and connection failures.

\textbf{Study Environment:}
Prior to deployment of studies, \appName automatically validates the \texttt{YAML} schema and verifies the reachability of linked VA systems. While syntactic and connection errors are caught systematically, \appName does not highlight semantic experimental design errors; researchers must manually verify logical linkages, such as cross-system task accomplishment notifications.

\subsubsection{Adaptability} \textit{What socio-technical (e.g., learnability) dimensions are considered?}
\appName is tailored for mixed-initiative (VA) researchers who have experience in developing mixed-initiative tools.
It provides a high architectural ceiling and a low experimental threshold. To ease building systems and studies, we provide tutorials and example implementations, allowing less experienced users to use \appName for their research.

\textbf{Infrastructure:}
Designing mixed-initiative systems requires full-stack programming proficiency to leverage our framework's capabilities. Nevertheless, \appName already handles complex structures, such as state synchronization and inter-agent communication.

\textbf{Study Environment:}
Once the core VA system is deployed, the \texttt{YAML}-based configuration provides a low-threshold entry point, allowing less technically versed researchers to design, modify, and execute complex multimodal studies without altering the underlying application code.

\section{User Evaluation: Expert Case Studies}
In the following, we present three case studies of projects built and evaluated using \appName, demonstrating its utility and ease of integration into both new and existing applications.
\subsection{Methodology \& Participants}
\paragraph{Methodology} The case studies were carried out over two weeks and started with an introduction session ($\sim$1 hour) in which we introduced the theoretical framework to the VA researchers, provided a tutorial for an example system, showed two example implementations of previous works (see \cref{sec:replicationWork}), and shared the codebase. In the following two weeks, participants implemented or retrofitted their systems in the backend and frontend, set up, and conducted pilot user studies. During this period, we kept in touch with participants to check on their progress and answer clarifying questions.
Because our evaluation focuses on validating the framework’s capability to streamline development and automate multi-modal data collection, the experts were instructed to conduct a focused pilot study with their system (three study sessions per expert tool). This scope was selected to thoroughly stress-test the functionality of \appName and validate the multi-modal logging pipeline and post-analysis capabilities across diverse mixed-initiative VA scenarios, without placing a prohibitive administrative burden on the experts participating in our case study. We concluded the case studies with semi-structured interviews to capture in-depth, formative feedback from experts on the utility and experience of using \appName (see Appendix \ref{appendix:interview}). Interviews were transcribed verbatim. 
Two authors iteratively reviewed the interview transcripts alongside the experts' notes taken during the two-week use of \appName. The feedback was grouped into key areas of interest: usability, workflows, development and integration effort, development strategies, and enhancement and feature requests.
We report these findings as descriptive observations.
\paragraph{Participants} We recruited three researchers ($P_{1-3}$) who conduct research in HCI or VA, currently designing or implementing mixed-initiative VA systems. They have rich experience (2-5 years) in design and development of multi-agent mixed-initiative systems, and have previously conducted at least three user studies. Participants were not paid, as they could use the built applications and pilot studies for their own research. This study was approved by ETH Zurich’s Ethics Commission (Project 26 ETHICS-244). 
\subsection{Developed Prototypes}
\label{sec:case_study_prototypes}
\begin{wrapfigure}[24]{r}{.50\textwidth}
    \vspace{-4.5em}
   \centering
    \includegraphics[width=.50\textwidth]{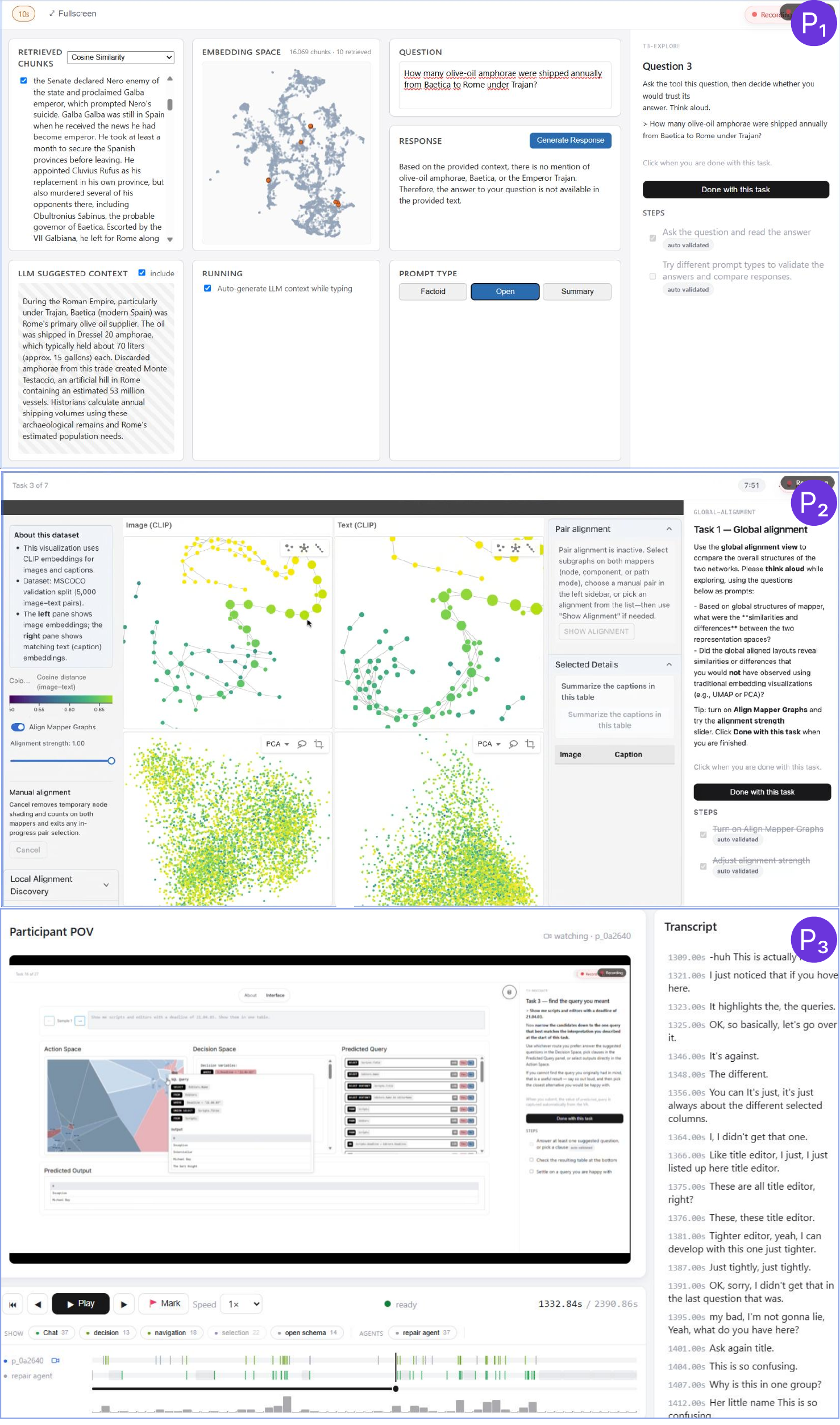}
    \vspace{-2.8em}
    \caption{Case study results implemented by the experts. UserRAG ($P_1$) and TopoAlign ($P_2$) are shown in deployed study mode, while PleaSQLarify ($P_3$) is shown in the Replay View of a study session. }
    \Description{}
    \label{fig:findings_modeltypes_per_year}
    \end{wrapfigure}
\subsubsection{UserRAG}
    UserRAG is an interactive mixed-initiative VA system where human and software agents collaborate to steer the retrieval and augmentation stages in a Retrieval Augmented Generation pipeline. Besides the generic retrieval, a software agent generates additional context on request of the human agent that can be further used to augment the retrieval. The researcher aims to evaluate their mixed-initiative approach by letting participants interact with the system and measure system usability and whether participants achieve better responses when the software agent augments the retrieval pipeline.

\subsubsection{TopoAlign}
TopoAlign~\cite{yan_2026_topoalign} is a topology-aware mixed-initiative VA framework that leverages mapper graph agents to help machine learning experts compare and analyze the structural alignment of neural network representations across different models, layers, or training conditions. 
The study evaluates whether mapper graph agents, compared to traditional embedding approaches, can uncover additional similarities or differences between two representation spaces and how machine learning experts interact with them.

\subsubsection{PleaSQLarify}
PleaSQLarify~\cite{chan_2026_pleasqlarify} is a mixed-initiative VA system designed for natural language database querying (text-to-SQL). It re-frames input ambiguity through the lens of pragmatic repair -- an incremental clarification process between human agent and the SQL software agent that helps align user intent with system interpretations. The authors' study aimed to evaluate pragmatic repair in text-to-SQL disambiguation by observing how agents navigate uncertain query spaces and assessing their overall control, interaction workflows, and system usability. Specifically, they investigated whether prioritized decision variables allow agents to efficiently and understandably filter the query space. Retrofitting the existing system with the \appName Infrastructure and replicating the study in the Study Environment allows assessing \appName's effectiveness along the full research pipeline.

\subsection{Expert and Developer Experience}
We refer to the experts as $P_1$ -- $P_3$, following the prototype order of \Cref{sec:case_study_prototypes}.

\subsubsection{General Feedback}
All experts characterized the agent-centric abstraction intuitive and quick to learn, since it reuses familiar engineering concepts: agents are ``very logical units'' like classes in object-oriented programming, and the World State maps to the model in model–view–controller ($P_1$). 
For participants, learning the underlying concepts of the \appName took from minutes ($P_2$) to an afternoon ($P_3$), supported by the introduction session, tutorial, and documentation ($P_{2,3}$). All three experts implemented their intended collaboration behaviors without reporting an expressiveness ceiling, as ``all the power of a programming language'' stays available, ``just placed at a different level'' ($P_1$) (\refdesigngoal{G5}{colorInfrastructure}, \refdesigngoal{G6}{colorInfrastructure}). 
Friction arose mainly in the declarative layer: getting used to the DSL grammar for agents, permissions, and study/task definitions took time (``time-consuming'' for $P_2$), and separating task and study files led to ``back and forth'' ($P_1$), though $P_1$ also noted ``great benefits in reusability of tasks.'' One expert needed our support to fix a defect when injecting a World State update within a study due to a variable naming error. 
\appName required no toolchain changes: experts used their familiar tools with the Python/TypeScript stack ($P_2$), browser-based deployment ($P_1$), and edited \texttt{YAML} files in their preferred editors. 

\subsubsection{Development Strategies}
All experts followed the same implementation sequence: agent decomposition and World State in the backend first, visual interface last. $P_1$ defined agents, their communication, and the World State, then built the frontend; $P_2$ and $P_3$ extracted main components from an existing system, defined agents and World State, then adjusted the frontend. The decomposition served as a design lens: $P_1$ noted they ``would not have looked at the problem on this agentic level'' otherwise and found it useful ``to look at the roles and abilities instead of just the functions'', which also made adding another agent mid-development straightforward (\refdesigngoal{G9}{colorStudyEnvironment}). Making permissions explicit was seen as helpful rather than overhead ($P_2$, \refdesigngoal{G7}{colorInfrastructure}), and routing all updates through the World State was considered ``better practice'' because all variables are centralized ($P_1$, \refdesigngoal{G8}{colorInfrastructure}). 

For study design, $P_2$ and $P_3$ derived additional study ideas from the theoretical concept and technical options offered by \appName.
 They first configured the core study setup and modalities to track, then defined the tasks.
As tasks and metrics reside outside the application (\refdesigngoal{G2}{colorGeneral}), $P_2$ explored different study setups once the system was deployed without the need to edit their VA application and captured additional metrics ``on the fly.''

\subsubsection{Development and Integration Effort}
Integration costs depended on the starting point: $P_2$ and $P_3$ retrofitted existing systems, while $P_1$ built on the \appName Infrastructure from scratch.
All experts observed a \emph{shift} and reduction in effort: implementing the mixed-initiative VA system itself is ``about the same'' as without \appName ($P_2$), with ``more focus on the agent behavior as \appName Infrastructure takes care of state synchronization, communication channels, etc.'' ($P_3$). In contrast, configuring, deploying, and iterating studies became much faster ($P_{1,2,3}$): complete setups were prepared within hours, and ablations created by copying tasks ($P_{1,3}$). Because these are self-assessments, we treat them as perceptions: the pilots covered the full pipeline (system implementation plus study definition, deployment, and execution) but are too small to quantify analysis savings.
$P_3$ offered the clearest comparison: their previous workflow for a user study spanned five sources—recordings, transcripts, notes, plus external questionnaire and consent forms—requiring about half a day per transcript pass plus extra hours revisiting recordings to contextualize user quotas. This was replaced by a single aligned record (\refdesigngoal{G1}{colorGeneral}, \refdesigngoal{G4}{colorGeneral}): sessions can be replayed to locate relevant events ($P_{2,3}$), and interlinking preserves the context of qualitative statements ($P_3$). $P_3$ emphasized this primarily as a quality gain: ``you can measure everything from the start.'' $P_1$ and $P_2$ appreciated that additional sensor streams only need to be enabled in the study setup (\refdesigngoal{G10}{colorStudyEnvironment}).
Contrary to our expectation (\Cref{sec:TDPScomplexity}), no expert found maintaining separate application and study models burdensome. Instead, $P_3$ noted that mapping application variables to study events adds a layer that ``can create confusion and errors'' and must be validated—an effort absorbed by a pilot run, which is good scientific practice~\cite{olson_2014_WaysOfKnowing}.

\subsection{Enhancement and Feature Requests}
Our expert participants suggested enhancements to \appName, including authoring support for the study DSL: a task-type selector with skeleton code ($P_1$), plus LLM agents to help generate task and study configurations ($P_2$). $P_2$ also requested video integration at the start of a study to replace live introductions, which we added in the latest version. For post-analysis, $P_1$ suggested filters to faster find specific study sessions, and $P_3$ recommended built-in verification of variable-to-event mappings to increase confidence in the data, a feature that we consider for future work. 
\section{Discussion}
\appName allows developers and researchers to build web-based mixed-initiative VA systems through a modular Infrastructure. Further, they can configure and deploy (multi-agent) user studies along their requirements that use the systems as stimuli. Our evaluation shows that the Infrastructure is flexible and modular enough to implement new or retrofit existing systems. Further, our expert case studies show that once implemented, researchers can configure and deploy their experiments and user studies with low effort. In addition, researchers like the diversity of data that can be collected through the Study Environment, and the optionality to add additional data streams from external sources, such as biometric sensors.

\subsection{Limitations \& Future Work}
While establishing a solid study framework, \appName’s current design operates under several inherent conceptual and technical limits.
    \paragraph{ \textbf{Adoption Effort and Cognitive Load}} While \appName provides an infrastructure for implementing mixed-initiative VA applications, it requires researchers and developers to align their application design and architecture with the theoretical concepts of \appName. While this improves standardization and application modularization, this might lead to additional effort and development overhead required to adapt to the provided infrastructure. Further, they need to maintain ``two separate mental models'' for the application state and the study setup, which might be cognitively demanding. Nevertheless, this burden comes to the benefit of being able to quickly set up and run studies with the implemented applications.
    \paragraph{ \textbf{Post-Analysis Enrichment}} Researchers can use the \appName Study Environment to collect a diverse set of data during a study session, also including sensor data. At the moment, data is visualized per data stream in the visual interface for session replay, but the interface does not allow joining data streams. Currently, e.g., this does not support visualizing eye tracking data, such as area of interest, on the captured screen replay. Such features would enrich the post-analysis, but also depend on individual requirements, which we defined to be out of scope for this work. Nevertheless, researchers can adapt the core implementation as the source code is open-sourced.   
    \paragraph{ \textbf{Desktop-First}} Currently, \appName is limited to desktop web-based mixed-initiative applications, and does not support VR/AR applications, nor is it tailored for mobile devices. Future work should address these modalities, as these are becoming more prominent over the years. 
    \paragraph{ \textbf{Browser Compatibility}} Some features, such as our heart rate collector, use the Web Bluetooth API for data collection, which is not supported by all browsers. A workaround is to use our generic Sensor API to push additional sensor data. We are aware of this limitation, and future work will address broader browser support for all integrated features.  
\subsection{Opportunities \& Next Steps}
Our efforts to provide a modular study framework for multi-agent, mixed-initiative VA systems reveal several directions for future work.

    \paragraph{\textbf{Sharing Systems and Studies}} \appName can be run by any individual locally or hosted. Researchers can use the framework to share their own research and make their results accessible and replicable by making use of \appName's modular structure.
    
    \paragraph{\textbf{Automated Testing of User Studies}} With the recent advancements in technology and the framework of \appName, we see great potential for researchers in reducing efforts for conducting pilot studies, as protocols like VACP~\cite{staehle_2026_vacp} might allow replacing human participants with AI agents to act on the stimulus in pilot studies, collect first data, and optimize the setup.
    \paragraph{\textbf{Flexible Study Design}}  \appName allows future studies to directly manipulate VA environment states and agents without heavy modification of the underlying VA system. E.g., an AI agent can be swapped to a human WoZ agent by adapting the agent permissions and roles, allowing for easier comparison of wizard agents and real software agents.
    \paragraph{\textbf{Human Behavior Cloning}} The large multi-variant data that can be collected through \appName allows for capturing human behavior and interactions. The data can be used for behavioral cloning into software agents and then testing the learned behavior in the same or a different environment, as systems and tasks can easily be changed.

\section{Conclusion}
In this work, we presented \appName, a multi-agent Study Environment for mixed-initiative VA systems. By providing a modular Infrastructure to build mixed-initiative VA systems and a Study Environment that eases the setup, configuration, and deployment of multi-agent experiments, \appName enables researchers to collect the relevant data for post-session analysis that captures insights on multi-agent collaboration and interaction in VA environments as well as user behavior data. This leverages evaluation of their implemented systems and approaches. We show that \appName's Infrastructure architecture allows building diverse mixed-initiative VA applications. Further, our expert case studies reveal that once a system is implemented on the \appName Infrastructure, researchers can configure, deploy, and run studies in a short time, providing a comprehensive set of data collection methods supporting mixing of qualitative and quantitative studies. To foster its use within the community, our framework is available as open source software at \faGithub:~\href{https://github.com/ETH-IVIA-Lab/MIVAIS}{github.com/ETH-IVIA-Lab/MIVAIS}.

\begin{acks}
We are grateful to our expert case study participants for their valuable time and feedback. 
\end{acks}

\section*{GenAI Usage Disclosure}
GenAI (Claude - Sonnet 5 and Opus 4.8) was partially used as a coding assistant during the implementation of \appName.
All system architecture design, expert case study protocols, analysis, and manuscript writing were conducted by the authors.

\bibliographystyle{ACM-Reference-Format}
\bibliography{references}

\newpage
\appendix

\section{Usage Scenario}
\label{appendix:usageSecenario}
To demonstrate our framework's practical value, we outline a typical workflow for designing, implementing, and evaluating a mixed-initiative VA application/approach.
The implementation pipeline then follows a multi-stage process:

\textbf{System Conceptualization:} The developer starts by outlining the core data model, relevant machine learning components, and drafting the VA interface. Crucially, this stage involves defining the agents and mapping their specific capabilities and action space to establish a comprehensive system prototype.

\textbf{Backend Orchestration:} The conceptual draft is implemented in the \appName backend. The developer defines the central World State by creating the environment schema, specifying state keys, and setting their initial values. Next, the agent configuration is declared by assigning unique identifiers, defining human and software agents, and setting strict observation and action permissions. The developer also configures event listeners, subscribes agents to communication channels, and sets hyperparameters for their behavior. Finally, the software agents' logic is implemented by extending the \appName base-agent class, operationalizing observations and actions, and structuring inter-agent communication.

\textbf{Frontend Integration:} The visual interface is wired through a dedicated context using the \appName Context Provider. The frontend communicates with the backend via the standardized \appName WebSocket API. At runtime, user identities are resolved and retrieved through the visual interface or URL parameters.

\textbf{Study Deployment and Post-Session Analysis:} After completing the system, the researcher uses \appName's Study Environment to configure the study sequence and specify analytical task parameters, then runs the setup in a controlled lab study. The framework then supports detailed post-session analysis, enabling systematic review of multimodal interaction trajectories and inter-agent communication, with direct annotation of recorded sessions for easy sharing of findings with stakeholders.

\textbf{Sharing Application and Study Setup for Replication:} Finally, the researcher can open-source the mixed-initiative VA application and share the study setup so the research community can replicate the work using the \appName Study Environment. Researchers can also share specific session replays via public links with reviewers and the community.

\section{Semi-Structured Interview Protocol} \label{appendix:interview}

\subsection{Introduction \& Consent} 
\begin{itemize}
    \item Briefly restate case study purpose, confidentiality, and recording consent.
    \item Confirmation of the expert’s role, background, and prior experience with Visual Analytics, mixed-initiative systems, or AI-human collaboration tooling.
\end{itemize}
\begin{enumerate}
    \item Could you briefly describe your background in Visual Analytics, HCI, or AI systems research/development?
\end{enumerate}

\subsection{Mixed-Initiative System-Building Experience} 
\begin{enumerate}[start=2]
    \item Can you walk me through the system you built with our \appName framework? What was its purpose, and what AI/human collaboration scenario did it support?
    \item What was your overall process for implementing it — what did you start with, and how did you proceed?
    \item Were there parts of the implementation that felt straightforward? Which parts, and why?
    \item Were there parts that felt difficult, confusing, or frustrating? Could you describe a specific moment where you got stuck?
    \item How did our \appName framework’s approach address state synchronization (between AI agents and human users), compared to how you would normally implement this from scratch?
\end{enumerate}

\subsection{Criteria-Based Assessment}
For each criterion below we ask the participant to give answers to the questions, and then probe the reasoning behind their response.

\textbf{Expressiveness}
\begin{enumerate}[start=7]
    \item To what extent did the \appName framework allow you to implement the interaction patterns and AI-human collaboration behaviors you envisioned?
    \item Were there any application and study designs or behaviors you wanted to implement but felt constrained or unable to express using \appName?
\end{enumerate}
\textbf{Creativity Support}
\begin{enumerate}[start=9]
    \item Did working with \appName influence the design decisions you made, positively or negatively? Did it open up ideas you wouldn't have otherwise considered, or did it push you toward more conventional solutions?
    \item Can you think of an instance where the \appName framework either enabled or limited a creative design choice?
\end{enumerate}
\textbf{Flexibility}
\begin{enumerate}[start=11]
\item How well did our software accommodate adapting or modifying your system and study design as your ideas evolved during development?
\end{enumerate}
\textbf{Efficiency}
\begin{enumerate}[start=12]
\item Compared to building a similar application and user study without \appName, how would you describe the development time and effort?
\item Which specific features, if any, saved you the most time or effort? Which, if any, slowed you down?
\end{enumerate}
\textbf{Usability}
\begin{enumerate}[start=14]
\item How would you describe your overall experience interacting with the \appName's API, and functionalities?
\end{enumerate}
\textbf{Learnability}
\begin{enumerate}[start=15]
\item How long did it take you to understand the core concepts and get a basic system and study running?
\item What aspects of \appName were easy to learn, and what aspects required more effort or external help (e.g., documentation, trial and error)? 
\end{enumerate}
\textbf{Integration into Existing Workflows}
\begin{enumerate}[start=17]
\item How well did our \appName framework fit into your existing development environment, tools, and practices (e.g., version control, testing, deployment)?
\item Would you be able to integrate the \appName framework into an existing project, or does it require building from scratch? Please explain.
\end{enumerate}
\subsection{Logging \& State-Recording Feature}
\begin{enumerate}[start=19]
\item Did the logged data and analysis meet your needs for analyzing application state and user/agent interactions? What, if anything, was missing or required manual post-processing?
\item Compared to how you would normally implement logging in a research prototype, how does this compare in terms of effort and data quality?
\end{enumerate}
\subsection{Comparative \& Reflective Questions}
\begin{enumerate}[start=21]
\item If you had to build the same system and study again, would you choose to use \appName again? Why or why not?
\item How does \appName compare to other tools or frameworks you've used for similar purposes (if any)?
\item Are there scenarios or application types you think the \appName framework would NOT be well-suited for?
\end{enumerate}
\subsection{Improvement Suggestions \& Closing}
\begin{enumerate}[start=24]
\item If you could change or add one thing to \appName, what would it be?
\item Is there anything about your experience that we haven't covered but that you think is important to share?
\end{enumerate}
\textbf{Closing:}
\begin{itemize}
    \item Thank you for participating in our expert case study. 
    \item Are there any additional questions on your side for our research team?
\end{itemize}

\section{\appName Domain Specific Language - Grammar}
\label{appendix:dsl_definition}
\subsection{Agent Registry - Role and Permission Definition}
\label{appendix:dsl_agent_definition}

Note on grammar notation: The ellipsis notation (\dots \Id{StructureName}) is used as a spread operator to denote the inline expansion and inclusion of all fields defined within the referenced \Id{StructureName}.

\noindent\textbf{AgentsConfig} = \\
$\left\{
\begin{array}{rcl}
\Id{agents} & ::= & \Id{AgentEntry}[] \\
\Id{users} & ::= & \Id{UserRole}[] \\
\Id{roles} & ::= & \{\, \Id{roleName} \rightarrow \Id{RoleDef} \,\} 
\end{array}
\right.$

\noindent\textbf{AgentEntry} = \\
$\left\{
\begin{array}{rcl}
\Id{agent\_id} & ::= & \Id{identifier} \\
\Id{class} & ::= & \Id{string} \\
\Id{enabled} & ::= & \Id{boolean} \\
\Id{role} & ::= & \Id{string} \\
  \dots\ \Id{PermissionFields} &&\\
\Id{params} & ::= & \{\, \Id{string} \rightarrow \Id{JSON} \,\} \\
\Id{observation} & ::= & \Id{Observation} \\
\Id{description} & ::= & \Id{string} 
\end{array}
\right.$

\noindent\textbf{Observation} = \\
$\left\{
\begin{array}{rcl}
\Id{strategy} & ::= & \Id{watch} \mid \Id{poll} \mid \Id{subscribe} \\
\Id{watches} & ::= & \Id{string}[]
\end{array}
\right.$

\noindent\textbf{UserRole} = \\
$\left\{
\begin{array}{rcl}
\Id{role} & ::= & \Id{identifier} \\
 \dots\ \Id{PermissionFields}& &
\end{array}
\right.$

\noindent\textbf{RoleDef} = \\
$\left\{
\begin{array}{rcl}
\Id{description} & ::= & \Id{string} \\
 \dots\ \Id{PermissionFields} & &
\end{array}
\right.$

\noindent\textbf{PermissionFields} = \\
$\left\{
\begin{array}{rcl}
\Id{can\_read} & ::= & \Id{string}[] \\
\Id{can\_write} & ::= & \Id{string}[] \\
\Id{communication\_bus\_publish\_topics} & ::= & \Id{string}[] \\
\Id{communication\_bus\_subscribe\_topics} & ::= & \Id{string}[]
\end{array}
\right.$

\noindent{\small
Semantics not expressible above: \Id{agent\_id} must be unique across the
list; \Id{class} must resolve in the application's agent-class map or the
entry is skipped with a warning}
\subsection{Study Configuration}
\label{appendix:dsl_study_config}
Note on grammar notation: The ellipsis notation (\dots \Id{StructureName}) is used as a spread operator to denote the inline expansion and inclusion of all fields defined within the referenced \Id{StructureName}.

\noindent\textbf{Study} = \\
$\left\{
\begin{array}{rcl}
\Id{id} & ::= & \Id{identifier} \\
\Id{name} & ::= & \Id{string} \\
\Id{version} & ::= & \Id{integer} \\
\Id{mode} & ::= & \Id{single-participant} \mid \Id{multi-participant} \\
\Id{participants\_required} & ::= & \Id{integer} \\
\Id{consent\_text\_md} & ::= & \Id{markdown} \\
\Id{va} & ::= & \Id{VASystems} \\
\Id{roles} & ::= & \Id{Role}[] \\
\Id{recording} & ::= & \Id{Recording} \\
\Id{ui} & ::= & \Id{UiConfig} \\
\Id{advance\_policy} & ::= & \Id{all} \mid \Id{first} \mid \Id{admin} \\
\Id{block\_order} & ::= & \Id{declared} \mid \Id{random} \mid \Id{round\_robin} \mid \Id{latin\_square} \\
\Id{block\_orders} & ::= & \Id{identifier}[\,][\,] \\
\Id{parameters} & ::= & \{\, \Id{string} \rightarrow \Id{JSON} \,\} \\
\Id{completion\_redirect\_url} & ::= & \Id{string} \\
\Id{completion\_code} & ::= & \Id{string} \\
\Id{metrics} & ::= & \Id{Metric}[] \\
\Id{blocks} & ::= & \Id{Block}[] \\
\Id{tasks} & ::= & \Id{TaskRef}[]
\end{array}
\right.$

\noindent{\small\Id{blocks} and \Id{tasks} are
mutually exclusive -- give one or the other; a bare \Id{tasks} list folds
into one implicit block.}

\noindent\textbf{VASystems} = \\
$\left\{
\begin{array}{rcl}
\Id{va\_systems} & ::= & \{\, \Id{identifier} \rightarrow \Id{VAConfig} \,\} \\
\Id{primary\_va\_system} & ::= & \Id{identifier}
\end{array}
\right.$

\noindent\textbf{VAConfig} = \\
$\left\{
\begin{array}{rcl}
\Id{system} & ::= & \Id{identifier} \\
\Id{external} & ::= & \Id{true} \\
\Id{iframe\_url} & ::= & \Id{string} \\
\Id{health\_check\_url} & ::= & \Id{string} \\
\Id{env} & ::= & \{\, \Id{string} \rightarrow \Id{string} \,\} \\
\Id{variants} & ::= & \{\, \Id{identifier} \rightarrow \Id{VAVariant} \,\} \\
\Id{recording\_dir\_override} & ::= & \Id{string} \\
\Id{token\_env} & ::= & \Id{string} \\
\Id{token\_param} & ::= & \Id{string} \\
\Id{room\_per\_session} & ::= & \Id{boolean} \\
\Id{room\_param} & ::= & \Id{string}
\end{array}
\right.$

\noindent\textbf{VAVariant} = \\
$\left\{
\begin{array}{rcl}
\Id{description} & ::= & \Id{string} \\
\Id{frontend\_id} & ::= & \Id{string} \\
\Id{backend\_id} & ::= & \Id{string} \\
\Id{spawn\_extra\_args} & ::= & \Id{string} \\
\Id{mivais\_config} & ::= & \{\, \Id{string} \rightarrow \Id{JSON} \,\}
\end{array}
\right.$

\noindent\textbf{Role} = \\
$\left\{
\begin{array}{rcl}
\Id{id} & ::= & \Id{identifier} \\
\Id{name} & ::= & \Id{string} \\
\Id{description} & ::= & \Id{string} \\
\Id{capacity} & ::= & \Id{integer} \\
\Id{selectable} & ::= & \Id{boolean} \\
\Id{color} & ::= & \Id{string} \\
\Id{agent\_id} & ::= & \Id{identifier} \\
\Id{wizard\_panel} & ::= & \Id{boolean}
\end{array}
\right.$

\noindent{\small\Id{agent\_id} names the \appName
Infrastructure role.
\Id{wizard\_panel} enables an optional Wizard-of-Oz control panel.}

\noindent\textbf{Recording} = \\
$\left\{
\begin{array}{rcl}
\Id{audio} & ::= & \Id{none} \mid \Id{per\_participant} \mid \Id{single\_mic} \\
\Id{video} & ::= & \Id{none} \mid \Id{screen} \\
\Id{transcribe} & ::= & \Id{boolean} \\
\Id{whisper\_model} & ::= & \Id{string} \\
\Id{whisper\_language} & ::= & \Id{string} \\
\Id{retention\_days} & ::= & \Id{integer} \\
\Id{biometric} & ::= & \Id{boolean} \\
\Id{external\_sensor} & ::= & \Id{boolean}
\end{array}
\right.$

\noindent{\small\Id{biometric}: BLE heart-rate strap
paired in the participant's own browser via Web Bluetooth.
\\\Id{external\_sensor}:
enables \texttt{POST /ingest/sensor-chunk} for non-browser sources (token-authenticated).}

\noindent\textbf{UiConfig} = \\
$\left\{
\begin{array}{rcl}
\Id{show\_progress\_bar} & ::= & \Id{boolean} \\
\Id{show\_timer} & ::= & \Id{boolean} \\
\Id{timer\_position} & ::= & \Id{top\_right} \mid \Id{top\_left} \mid \Id{sidebar} \mid \Id{hidden} \\
\Id{task\_description\_position} & ::= & \Id{top} \mid \Id{left} \mid \Id{overlay} \\
\Id{task\_steps\_visible} & ::= & \Id{boolean} \\
\Id{task\_steps\_position} & ::= & \Id{right} \mid \Id{left} \mid \Id{hidden} \\
\Id{show\_presence} & ::= & \Id{boolean} \\
\Id{theme} & ::= & \Id{light} \mid \Id{dark} \mid \Id{auto}
\end{array}
\right.$

\noindent\textbf{Block} = \\
$\left\{
\begin{array}{rcl}
\Id{id} & ::= & \Id{identifier} \\
\Id{name} & ::= & \Id{string} \\
\Id{description\_md} & ::= & \Id{markdown} \\
\Id{randomize\_within} & ::= & \Id{boolean} \\
\Id{tasks} & ::= & \Id{TaskRef}[]
\end{array}
\right.$

\noindent\textbf{TaskRef} $::=$ \quad
$\Id{identifier} \mid \Id{RefWithOverrides} \mid \Id{Task}$

\noindent{\small A bare \Id{identifier} resolves to
\texttt{tasks/<identifier>.yaml}; \Id{RefWithOverrides} is a shared fragment
with a shallow merge; \Id{Task} is an inline task object.}

\noindent\textbf{RefWithOverrides} = \\
$\left\{
\begin{array}{rcl}
\Id{\$ref} & ::= & \Id{string} \\
 \dots\ \Id{TaskFieldOverride} & &
\end{array}
\right.$

\noindent\textbf{Metric} = \\
$\left\{
\begin{array}{rcl}
\Id{id} & ::= & \Id{identifier} \\
\Id{label} & ::= & \Id{string} \\
\Id{kind} & ::= & \Id{count} \mid \Id{rate} \mid \Id{latency} \mid \Id{ratio} \mid \Id{python} \\
\Id{task} & ::= & \Id{identifier} \\
 \dots\ \Id{MetricKindFields} & &
\end{array}
\right.$

\noindent{\small Metric allows to measure additional (behavioral) metrics when agents interact with a VA system.}

\noindent\textbf{MetricKindFields} $::=$ \quad
$\Id{MatchFields} \mid \Id{LatencyFields} \mid \Id{RatioFields} \mid \Id{PythonFields}$

\noindent\textbf{MatchFields} = \\
$\left\{
\begin{array}{rcl}
\Id{match} & ::= & \Id{EventMatch}
\end{array}
\right.$
\quad{\footnotesize\color{mutedgray}(kind: count \textbar\ rate)}

\noindent\textbf{LatencyFields} = \\
$\left\{
\begin{array}{rcl}
\Id{from} & ::= & \Id{EventMatch} \\
\Id{to} & ::= & \Id{EventMatch} \\
\Id{aggregate} & ::= & \Id{mean} \mid \Id{median} \mid \Id{min} \mid \Id{max} \\
\Id{first\_only} & ::= & \Id{boolean}
\end{array}
\right.$
\quad{\footnotesize\color{mutedgray}(kind: latency)}

\noindent\textbf{RatioFields} = \\
$\left\{
\begin{array}{rcl}
\Id{numerator} & ::= & \Id{EventMatch} \\
\Id{denominator} & ::= & \Id{EventMatch}
\end{array}
\right.$
\quad{\footnotesize\color{mutedgray}(kind: ratio; value = n/(n+d))}

\noindent\textbf{PythonFields} = \\
$\left\{
\begin{array}{rcl}
\Id{module} & ::= & \Id{string} \\
\Id{params} & ::= & \{\, \Id{string} \rightarrow \Id{JSON} \,\}
\end{array}
\right.$
\quad{\footnotesize\color{mutedgray}(kind: python script)}

\noindent\textbf{EventMatch} = \\
$\left\{
\begin{array}{rcl}
\Id{event\_type} & ::= & \Id{string} \\
\Id{source} & ::= & \Id{vasystem} \mid \Id{studio} \\
\Id{meta} & ::= & \{\, \Id{string} \rightarrow \Id{JSON} \,\}
\end{array}
\right.$

\subsection{Task Configuration}
\label{appendix:dsl_task_config}

Note on grammar notation: The ellipsis notation (\dots \Id{StructureName}) is used as a spread operator to denote the inline expansion and inclusion of all fields defined within the referenced \Id{StructureName}.

\noindent\textbf{Task} = \\
$\left\{
\begin{array}{rcl}
\Id{id} & ::= & \Id{identifier} \\
\Id{type} & ::= & \Id{TaskType} \\
 \dots\ \Id{BaseFields} & &\\ 
 \dots\ \Id{TypeFields} & &
\end{array}
\right.$

\noindent{\small\Id{id} defaults to the filename.}

$\begin{array}{rcl}
\textbf{TaskType}& ::= & \Id{info\_screen}_\theta \mid \Id{single\_choice}_\theta \mid \Id{multi\_choice}_\theta \mid \Id{likert}_\theta \\
& \mid & \Id{slider}_\theta \mid \Id{number\_input}_\theta \mid \Id{free\_text}_\theta \mid \Id{VA\_interaction}_\theta
\end{array}$

\noindent{\small$\theta$ marks that task type
carries its own extra fields, listed under ``Per-type fields'' below.}

\noindent\textbf{BaseFields} = \\
$\left\{
\begin{array}{rcl}
\Id{prompt\_md} & ::= & \Id{markdown} \\
\Id{prompt\_md\_by\_role} & ::= & \{\, \Id{identifier} \rightarrow \Id{markdown} \,\} \\
\Id{timer\_seconds} & ::= & \Id{integer} \\
\Id{optional} & ::= & \Id{boolean} \\
\Id{tags} & ::= & \Id{string}[] \\
\Id{task\_steps} & ::= & \Id{TaskStep}[] \\
\Id{ground\_truth} & ::= & \Id{GroundTruth} \\
\Id{with\_va} & ::= & \Id{boolean} \\
\Id{va\_system} & ::= & \Id{identifier} \\
\Id{world\_state} & ::= & \Id{WorldState} \\
\Id{advance} & ::= & \Id{AdvanceOverride}
\end{array}
\right.$

\noindent\textbf{TaskStep} = \\
$\left\{
\begin{array}{rcl}
\Id{id} & ::= & \Id{identifier} \\
\Id{label} & ::= & \Id{string} \\
\Id{required} & ::= & \Id{boolean} \\
\Id{mode} & ::= & \Id{manual} \mid \Id{auto} \mid \Id{either} \\
\Id{auto\_check\_on} & ::= & \Id{AutoCheck}
\end{array}
\right.$

\noindent\textbf{AutoCheck} = \\
$\left\{
\begin{array}{rcl}
\Id{event\_type} & ::= & \Id{string} \\
\Id{count} & ::= & \Id{integer} \\
\Id{meta} & ::= & \{\, \Id{string} \rightarrow \Id{JSON} \,\}
\end{array}
\right.$

\noindent\textbf{GroundTruth} = \\
$\left\{
\begin{array}{rcl}
\Id{type} & ::= & \Id{exact} \mid \Id{set\_match} \mid \Id{ordered\_match} \mid \Id{regex} \mid \Id{custom} \\
\Id{value} & ::= & \Id{JSON} \\
\Id{partial\_credit} & ::= & \Id{boolean}
\end{array}
\right.$

\noindent\textbf{WorldState} = \\
$\left\{
\begin{array}{rcl}
\Id{mode} & ::= & \Id{continue} \mid \Id{replace} \\
\Id{initial} & ::= & \{\, \Id{string} \rightarrow \Id{JSON} \,\} \\
\Id{set} & ::= & \{\, \Id{string} \rightarrow \Id{JSON} \,\} \\
\Id{by\_role} & ::= & \{\, \Id{identifier} \rightarrow \Id{WorldState} \,\}
\end{array}
\right.$

\noindent{\small\Id{initial} is the full replace-mode scenario. \Id{set} is a delta on top of the current state. 
\\Both are written through the Study Gateway at task start.}

\noindent\textbf{AdvanceOverride} = \\
$\left\{
\begin{array}{rcl}
\Id{policy} & ::= & \Id{all} \mid \Id{first} \mid \Id{admin} \\
\Id{timeout\_seconds} & ::= & \Id{integer}
\end{array}
\right.$

\noindent\textbf{InfoFields} = \\
$\left\{
\begin{array}{rcl}
\Id{body\_md} & ::= & \Id{markdown} \\
\Id{continue\_label} & ::= & \Id{string} \\
\Id{min\_view\_seconds} & ::= & \Id{integer}
\end{array}
\right.$

\noindent\textbf{SingleChoiceFields} = \\
$\left\{
\begin{array}{rcl}
\Id{options} & ::= & \Id{Option}[]
\end{array}
\right.$

\noindent\textbf{MultiChoiceFields} = \\
$\left\{
\begin{array}{rcl}
\Id{options} & ::= & \Id{Option}[] \\
\Id{min\_selected} & ::= & \Id{integer} \\
\Id{max\_selected} & ::= & \Id{integer}
\end{array}
\right.$

\noindent\textbf{Option} = \\
$\left\{
\begin{array}{rcl}
\Id{id} & ::= & \Id{identifier} \\
\Id{label} & ::= & \Id{string}
\end{array}
\right.$

\noindent\textbf{LikertFields} = \\
$\left\{
\begin{array}{rcl}
\Id{scale} & ::= & \Id{LikertScale} \\
\Id{items} & ::= & \Id{LikertItem}[] \\
\Id{randomize\_items} & ::= & \Id{boolean}
\end{array}
\right.$

\noindent\textbf{LikertScale} = \\
$\left\{
\begin{array}{rcl}
\Id{min} & ::= & \Id{integer} \\
\Id{max} & ::= & \Id{integer} \\
\Id{min\_label} & ::= & \Id{string} \\
\Id{max\_label} & ::= & \Id{string} \\
\Id{step} & ::= & \Id{integer}
\end{array}
\right.$

\noindent\textbf{LikertItem} = \\
$\left\{
\begin{array}{rcl}
\Id{id} & ::= & \Id{identifier} \\
\Id{label} & ::= & \Id{string} \\
\Id{description} & ::= & \Id{string} \\
\Id{reverse} & ::= & \Id{boolean}
\end{array}
\right.$

\noindent\textbf{SliderFields} = \\
$\left\{
\begin{array}{rcl}
\Id{min} & ::= & \Id{number} \\
\Id{max} & ::= & \Id{number} \\
\Id{step} & ::= & \Id{number} \\
\Id{min\_label} & ::= & \Id{string} \\
\Id{max\_label} & ::= & \Id{string} \\
\Id{unit} & ::= & \Id{string} \\
\Id{default\_value} & ::= & \Id{number} \\
\Id{show\_value} & ::= & \Id{boolean}
\end{array}
\right.$

\noindent\textbf{NumberFields} = \\
$\left\{
\begin{array}{rcl}
\Id{min} & ::= & \Id{number} \\
\Id{max} & ::= & \Id{number} \\
\Id{step} & ::= & \Id{number} \\
\Id{unit} & ::= & \Id{string} \\
\Id{placeholder} & ::= & \Id{string}
\end{array}
\right.$

\noindent\textbf{FreeTextFields} = \\
$\left\{
\begin{array}{rcl}
\Id{min\_chars} & ::= & \Id{integer} \\
\Id{max\_chars} & ::= & \Id{integer} \\
\Id{placeholder} & ::= & \Id{string} \\
\Id{rows} & ::= & \Id{integer}
\end{array}
\right.$

\noindent\textbf{VAInteractionFields} = \\
$\left\{
\begin{array}{rcl}
\Id{variant} & ::= & \Id{identifier} \\
\Id{parameters} & ::= & \{\, \Id{string} \rightarrow \Id{JSON} \,\} \\
\Id{mivais\_config} & ::= & \{\, \Id{string} \rightarrow \Id{JSON} \,\} \\
\Id{answer} & ::= & \Id{VAAnswer}
\end{array}
\right.$
\quad{\footnotesize\color{mutedgray}(\Id{variant} default "default")}

\medskip
\noindent\textbf{VAAnswer} = \\
$\left\{
\begin{array}{rcl}
\Id{type} & ::= & \Id{capture\_state} \mid \Id{text} \mid \Id{choice} \mid \Id{none} \\
\Id{world\_state\_key} & ::= & \Id{string} \\
\Id{required} & ::= & \Id{boolean}
\end{array}
\right.$

\noindent{\small
Semantics not expressible above: \Id{ground\_truth} applies only to
answerable types.}

\end{document}